\documentclass[aps,prb,showpacs,citeautoscript,superscriptaddress,twocolumn]{revtex4-1}

\usepackage{graphicx}
\usepackage{amsmath}
\usepackage{amssymb}
\usepackage{color}
\usepackage[utf8]{inputenc}
\usepackage{braket}
\usepackage{xfrac}

\newcommand{\bea}{\begin{eqnarray*}}
\newcommand{\eea}{\end{eqnarray*}}
\newcommand{\bne}{\begin{equation*}}
\newcommand{\ede}{\end{equation*}}

\newcommand{\bnen}{\begin{equation}}
\newcommand{\eden}{\end{equation}}
\newcommand{\bean}{\begin{eqnarray}}
\newcommand{\eean}{\end{eqnarray}}
\newcommand{\bnsn}{\begin{subequations}}
\newcommand{\edsn}{\end{subequations}}

\newcommand{\bna}{\begin{array}}
\newcommand{\eda}{\end{array}}
\newcommand{\bnm}{\begin{enumerate}}
\newcommand{\edm}{\end{enumerate}}
\newcommand{\bni}{\begin{itemize}}
\newcommand{\edi}{\end{itemize}}

\renewcommand{\vec}[1]{\text{\boldmath{$ #1 $}}}

\renewcommand{\vec}[1]{\text{\boldmath{$ #1 $}}}
\newcommand{\trm}{\textrm}
\newcommand{\mrm}{\mathrm}

\begin{document}

\title{Valley relaxation in graphene due to charged impurities}

\author{P\'eter Boross}
\affiliation{Institute of Physics, E\"otv\"os University, Budapest, Hungary}

\author{Andr\'as P\'alyi}
\affiliation{Institute of Physics, E\"otv\"os University, Budapest, Hungary}
\affiliation{MTA-BME Condensed Matter Research Group, Budapest University of Technology and Economics, Budapest, Hungary}

\date{\today}

\begin{abstract}
Monolayer graphene is an example of materials with multi-valley electronic structure. 
In such materials, the valley index is being considered as an information carrier.
Consequently, relaxation mechanisms leading to loss of valley information are of interest. 
Here, we calculate the rate of valley relaxation induced by charged impurities in graphene.
A special model of graphene is applied, where the $p_z$ orbitals are two-dimensional Gaussian functions, with a spatial extension characterised by an effective Bohr radius $a_\textrm{eB}$. 
We obtain the valley relaxation rate by solving the Boltzmann equation, for the case of noninteracting electrons, as well as for the case when the impurity potential is screened due to electron-electron interaction. 
For the latter case, we take into account local-field effects and evaluate the dielectric matrix in the random phase approximation. 
Our main findings:
(i) The valley relaxation rate is proportional to the electronic density of states at the Fermi energy.
(ii) Charged impurities located in the close vicinity of the graphene plane, at distance $d \lesssim 0.3\,\textrm{\AA}$, are much more efficient in inducing valley relaxation than those farther away, the effect of the latter being suppressed exponentially with increasing graphene-impurity distance $d$.
(iii) Both in the absence and in the presence of electron-electron interaction, the valley relaxation rate shows pronounced dependence on the effective Bohr radius $a_\textrm{eB}$. The trends are different in the two cases: 
in the absence (presence) of screening, the valley relaxation rate decreases (increases) for increasing effective Bohr radius. 
This last result highlights that a quantitative calculation of the valley relaxation rate should incorporate electron-electron interactions as well as an accurate knowledge of the electronic wave functions on the atomic length scale.
\end{abstract}

\maketitle

\section{Introduction}
Certain crystalline solids, such as monolayer and bilayer graphene, carbon nanotubes, transition-metal dichalcogenides, silicon, and diamond, possess multi-valley electronic structure. 
Recently, ways to  control and measure the valley degree of freedom (or \emph{valley index}, for short) in these materials have been proposed\cite{Rycerz,DiXiaoPRL2007,WangYaoPRB2008,Gunlycke,WangKongTse,WenYuShanPRB2015,Golub,WehlingPRB2015} and
tested experimentally\cite{Zhu,Cao,Mak,Zeng,Isberg,Laird-nn,GorbachevScience2014,MakScience2014}.
The valley index is also being actively considered as a carrier of quantum information \cite{Recher,Palyi-valley-resonance,WuPRB2011,WuPRB2012,WuPRB2013,CulcerPRL,Laird-nn,Szechenyi-maximalrabi,KormanyosPRX2014,GuiBinLiuNJP2014,Rohling-njp,Rohling-prl}.

The valley index of an electron is linked to its crystal momentum. 
The crystal momentum is changed upon scattering, and consequently, the electron can be moved between different valleys by scattering processes (\emph{intervalley scattering}) in an uncontrolled, random fashion. 
Therefore, such scattering processes lead to the loss of information encoded in the valley index.

In this work, we theoretically study how (classical) valley information, encoded in an ensemble of electrons, is lost due to scattering processes. 
We focus on a specific multi-valley material, monolayer graphene\cite{Geim-rise,CastroNeto-rmp}, because of its relatively simple band structure and  widespread experimental availability. 
In particular, we consider  intravalley and intervalley scattering of graphene’s electrons off nearby charged impurities (Coulomb scattering), and calculate the corresponding \emph{valley relaxation time}, that is, the time scale characterizing the loss of valley information.

Coulomb scattering is a well-studied mechanism as a determinant of the electrical conductivity of graphene\cite{CastroNeto-rmp,AndoJPSJ,HwangPRB-transport}, but to our knowledge, its role in intervalley scattering has not been studied in detail. 
In fact,  the Coulomb potential of a charged impurity is expected to be less efficient in inducing intervalley scattering than in inducing intravalley scattering, since the Fourier spectrum of the Coulomb potential is peaked around small wave numbers.
Here, we set out to go beyond that qualitative argument by quantifying the efficiency of Coulomb scattering for valley relaxation.

At present, the understanding of valley relaxation processes is rather limited; the various mechanisms, such as electron-phonon, electron-impurity and electron-electron scattering, and their material-specific details, are yet to be systematically investigated.
Note, however, that recent studies have started to elucidate various aspects of valley relaxation and decoherence in two-dimensional (2D)
transition-metal dichalcogenides\cite{Song,Ochoa,CongMaiPRB2014} and graphene\cite{Pachoud,Braginsky}, as well as carbon-based\cite{WuPRB2011,Csiszar} and silicon quantum dots\cite{Tahan,Yang}.

To describe valley relaxation in graphene due to charged impurities, 
we use a special model of graphene's electrons, 
in which the $p_z$ orbitals are described by 2D
Gaussian functions, with a spatial extension characterised by 
an effective Bohr radius $a_\trm{eB}$. 
We obtain the valley relaxation rate $\Gamma_\trm{v}$
by solving
the corresponding Boltzmann equation, 
for the case of noninteracting electrons,
as well as for the case when the impurity potential is
screened due to electron-electron interaction. 
For the latter case, we calculate the screened impurity potential by taking into account local-field effects
and evaluating the dielectric matrix in the random phase approximation (RPA)\cite{AdlerPR,WiserPR,Schilfgaarde,Tudorovskiy}. 

Our main findings are as follows. 
(i) The valley relaxation rate is proportional to the electronic density
of states at the Fermi energy.
(ii) Charged impurities located in the close vicinity of the
graphene plane, at distance $d \lesssim 0.3\,\trm{\AA}$, are much more efficient in inducing
valley relaxation than those farther away, the effect of the latter being
suppressed exponentially with increasing graphene-impurity distance $d$.
(iii) Both in the absence and in the presence of electron-electron interaction,
the valley relaxation rate shows pronounced dependence 
on the effective Bohr radius $a_\trm{eB}$. 
Remarkably, the trends are different in the two cases: 
in the absence (presence) of screening, the valley relaxation rate
decreases (increases) for increasing effective Bohr radius. 
This last result highlights that a quantitative calculation of
the valley relaxation rate should incorporate electron-electron interactions
as well as an accurate knowledge of the electronic
wave functions on the atomic length scale. 

It should be emphasized that intervalley scattering has consequences beyond inducing valley relaxation: it has its fingerprints on the magnetoconductivity  as well as on inelastic light scattering, i.e., the Raman spectrum. In graphene, the quantum correction to the conductivity is influenced by elastic intervalley scattering processes\cite{Suzuura,McCannPRL}: for weak (strong) intervalley scattering, the correction to the conductivity is positive (negative), corresponding to weak antilocalisation (weak localisation). In experiments, a negative correction can be observed which is attributed to a significant intervalley scattering rate\cite{MorozovPRL,WuPRL2007,TikhonenkoPRL2008}. In Raman spectra, the D peak intensity  increases with increasing intervalley scattering.\cite{MalardPR2009} Furthermore, the intervalley scattering rate can be monitored in real space via spatially resolved Raman spectroscopy\cite{GrafNanoLett2007}, revealing that the sample boundary can be a strong source of intervalley scattering.

\section{Preliminaries}\label{prelim}
\begin{figure}
	\includegraphics[width=1\columnwidth]{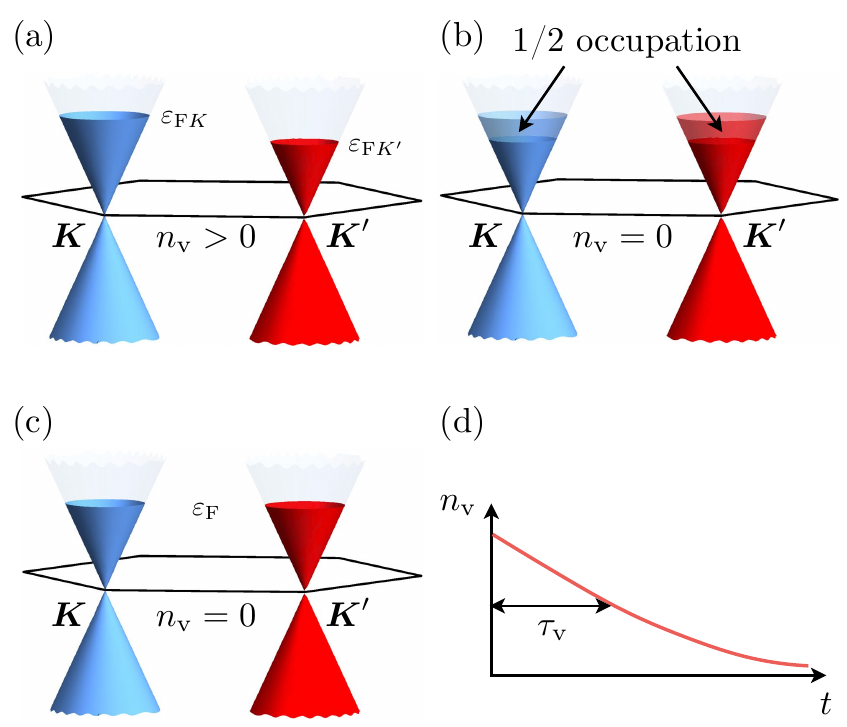}
	\caption{\label{fig1}
	Evolution of a valley-polarized initial state due to scattering processes. 
	(a) Initial state with a finite valley polarization $n_\trm{v}$. 
	(b) Non-equilibrium valley-unpolarized state, which appears during the relaxation process if the elastic intervalley scattering is faster than the inelastic processes. 
	(c) Thermal equilibrium, reached from (a) or (b), due to inelastic scattering processes. 
	(d) Schematic representation of the time evolution of valley polarization. 
	The characteristic time scale of the decay is the valley relaxation time $\tau_\trm{v}$. 	}
\end{figure}

We define the valley polarization $n_\trm{v}$ as the imbalance of the electronic populations in the two valleys ($n_{K}$ and $n_{K'}$),  i.e.,  $n_\trm{v}=n_{K}-n_{K'} $. Our aim is to describe the dynamics of the valley-polarized initial state shown in Fig.~\ref{fig1}a, under the influence of impurity scattering. It is expected that impurity scattering transfers electrons from one valley to the other (intervalley scattering), and therefore leads to the decay of the valley polarization with time, see Fig.~\ref{fig1}d. The task is to quantify the time evolution of this decay.

The qualitative nature of the dynamics of the electron distribution depends strongly on the relative time scales of elastic and inelastic scattering process. 
In this work, we will describe how elastic intervalley scattering caused by static impurities contributes to the decay of valley polarization; hence, we will disregard inelastic processes. 
Under the assumption that inelastic processes are absent, the valley-polarized initial state shown in Fig.~\ref{fig1}a evolves to the non-equilibrium, but valley unpolarized state depicted in Fig.~\ref{fig1}b.
If inelastic intravalley transitions are present, but they are slow compared to the elastic intervalley processes, then they reinforce thermal equilibrium (Fig.~\ref{fig1}c) after the intermediate state in Fig.~\ref{fig1}b is reached.  
However, if inelastic intravalley transitions are faster than elastic intervalley processess, then the initial state of Fig.~\ref{fig1}a evolves directly toward thermal equilibrium (Fig.~\ref{fig1}c).
Even though we do not incorporate inelastic processes in our model below, we expect that our treatment provides an accurate description of the valley relaxation time in all cases discussed above.

\begin{figure}
	\includegraphics[width=1\columnwidth]{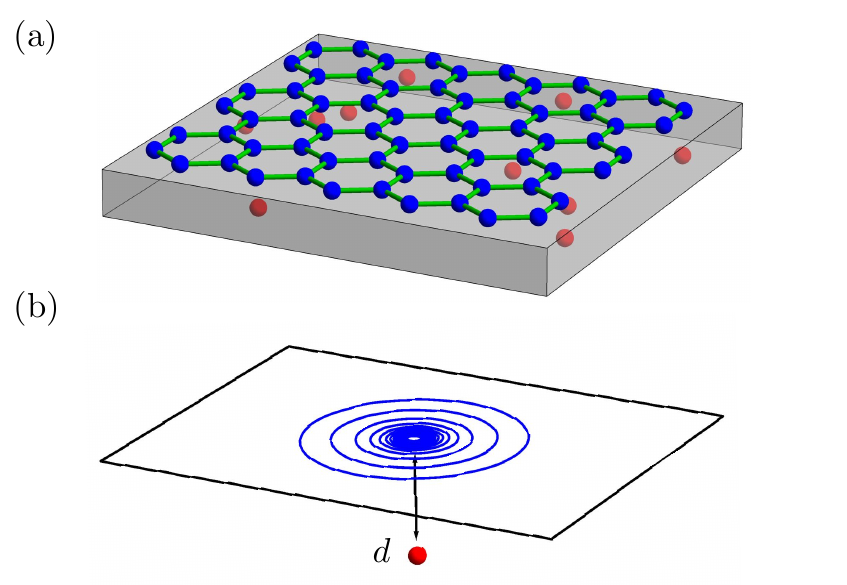}
	\caption{Charged impurities as a source of valley relaxation. (a) Random spatial arrangement of charged impurities (red spheres) in the substrate (gray) supporting the graphene sheet. (b) Schematic representation of the impurity-induced Coulomb potential in the plane of the graphene sheet.}
	\label{fig2}
\end{figure}

It is customary to distinguish short-range and long-range impurities. The short-range label usually refers to crystallographic defects, charge-neutral adatoms, etc. Long-range refers to charged scatterers that induce a long-range Coulomb potential for the mobile electrons. These impurities might be located, e.g., in the substrate supporting the graphene sheet, as shown in Fig.~\ref{fig2}a. For the present paper, we consider a model where the relaxation of valley polarization is due to charged impurities that are randomly positioned in a plane at a given distance $d$ from the graphene sheet (see Fig.~\ref{fig2}b). (Generalization of our methods to other impurity types and spatial distributions is probably straightforward.)

Figure \ref{fig2}b shows a single charged impurity located at a distance $d$ from the graphene sheet. Assuming this is a negatively charged impurity with charge $-e$, it creates a repulsive potential energy landscape $V_{\trm{i}}(\vec r; d$) for the delocalized electrons in graphene;
here $\vec r  =(x,y)$ is the position vector in the graphene plane. 
If the impurity is located on the $z$ axis of the reference frame, then
the  2D Fourier transform of the potential 
$V_{\trm i}$ reads 
\bnen
V_{\trm{i}}\left(\vec{q};d\right) =\frac{2\pi e_0^2}{q}\mrm{e}^{-qd},
\label{eq:unscreenedpotential}
\eden
where we use $e_0^2=e^2/4\pi\epsilon_0$.
Here, $e$ is the magnitude of the electron charge and $\epsilon_0$ is the
vacuum permittivity. 

\section{Model}

\subsection{Boltzmann equation and the valley relaxation time}

In this section, we calculate the valley relaxation rate due to elastic electron-impurity scattering in the framework of Boltzmann theory. 
We find that the
valley polarization $n_{\trm v}$ decays exponentially with time, 
and obtain a simple relation [Eq.~\eqref{eq:valleyrelaxtime1}] between the corresponding rate,
i.e., the \emph{valley relaxation rate} and the
momentum-dependent intervalley scattering rates.

For clarity, we assume that the valence band is filled, and the distribution function characterizing the occupation of bulk conduction-band states is denoted by $f_{\vec k}$, i.e., it is assumed to be independent of position. 
Then the Boltzmann equation reads:
\bnen
\label{eq:fBoltzmann}
\frac{\partial f_{\vec k}}{\partial t} = 
\sum_{\vec k'} W_{\vec k \vec k'} (f_{\vec k'} - f_{\vec k}),
\eden
where $W_{\vec k \vec k'}$ is the impurity-induced transition rate from state $ \vec{k} $ to state $ \vec{k}' $. Here we assumed detailed balance $W_{\vec k \vec k'} = W_{\vec k' \vec k}$, which is reasonable as the scattering rates will be evaluated using Fermi's Golden Rule:
\bnen
\label{eq:FGR}
W_{\vec{kk}'} = \frac{2\pi}{\hbar} \overline{\left|V_{\trm{c}\vec{k},\trm{c}\vec{k}'}\right|^2}\delta\left(\varepsilon_{\trm{c}\vec{k}}-\varepsilon_{\trm{c}\vec{k}'}\right).
\eden
Here, 
$V_{\trm{c}\vec{k},\trm{c}\vec{k}'}=\braket{\trm{c}\vec{k}|V|\trm{c}\vec{k}'}$ is the
matrix element of the impurity potential 
$V(\vec r;d) = \sum_{j=1}^{N_{\trm{i}}} V_{\trm{i}}(\vec r - \vec r_j;d)$ 
with 
the Bloch-type energy eigenstates $\ket{\trm{c}\vec k}$ of the conduction band
(see below).
The impurity potential $V(\vec r;d)$ 
is the electric potential energy created by the impurity ensemble with $N_{\trm i}$ impurities.
The overline denotes disorder average, with the impurity positions
$\vec r_i$ assumed to be independently and homogeneously distributed.
$\varepsilon_{\trm{c}\vec k}$ denotes the energy of the conduction-band 
state at momentum $\vec k$. 

The solution of the Boltzmann equation will be facilitated by the fact that in graphene, the low-energy excitations have an approximately conical
(linear and isotropic) dispersion relation 
around the Dirac points \cite{WallacePR,CastroNeto-rmp}.
Here, we consider a sample which has a Fermi energy $\varepsilon_\trm{F}$ that
is small compared to the energy width 
of the $\pi$ band, the latter being approximately $16.8\,\trm{eV}$. 
This has the following implications. 
First, the dispersion is approximately conical (see Fig.~\ref{fig1}) at energy
$\varepsilon_\trm{F}$, and
the two Fermi lines are approximately circles with radii $\kappa_\trm{F} = 
\varepsilon_\trm{F} / \hbar v_\trm{F}$, where $v_\trm{F} \approx 9\times10^5\,\trm{m/s}$ is
the Fermi velocity. 
Second, the crystal momenta $\vec k$ of the electronic states 
participating in the valley relaxation process are close to either $\vec K$
or $\vec K'$, hence we can uniquely relabel their momentum as 
$\vec k = \vec K + \vec \kappa$
or $\vec k = \vec K' + \vec \kappa$, respectively,
with $\kappa \equiv |\vec \kappa| \ll K$.

Let us briefly and qualitatively discuss the energy range where the
conical (linear and isotropic) approximation of the electronic dispersion is valid. 
We have checked that the error of the conical approximation 
[see Eq.~\eqref{eq:lindisp}]
with respect to the tight-binding dispersion [see Eq.~\eqref{eq:disp}] 
is at most $10\%$ 
for wave vectors fulfilling $\kappa < \kappa_\trm{c} =0.316/a_\trm{CC}= 2.23\times10^{9}\,\trm{m}^{-1}$,
i.e., within this wave-vector range 
the relations 
$
0.9 \, \varepsilon^{\trm{(lin)}}_{n,\vec K+\vec \kappa}
< 
 \varepsilon_{n,\vec K+\vec \kappa}
< 
1.1 \, \varepsilon^{\trm{(lin)}}_{n,\vec K+\vec \kappa}
$ are fulfilled.
The Fermi energy corresponding to $\kappa_\trm{c}$ is 
$\hbar v_\trm{F} \kappa_\trm{c} \approx 1.33\,\trm{eV}$.
Therefore, it is reasonable to use the conical approximation 
of the electronic dispersion as long as 
\bnen
\label{eq:lincond}
\varepsilon_\trm{F}\lesssim 1.33\,\trm{eV}.
\eden

The Boltzmann-type description of carrier dynamics
in graphene in the presence of Coulomb scatterers 
is expected to be valid if the carrier density exceeds 
a threshold set by the 
graphene-impurity distance $d$
and the
impurity density $n_\textrm{i} = 
N_\textrm{i}/A$, 
where $A$ is the sample area.
The theoretical and experimental grounds of this expectation
are summarized in, e.g.,
Sec. III. A of Ref.~\onlinecite{DasSarma-graphenereview} 
and in Ref.~\onlinecite{HwangPRB-transport}.
There are strong indications that 
the Boltzmann-type description is invalid for 
very low carrier densities, when the Fermi energy is in the close vicinity of the 
Dirac points; see, e.~g., Refs.~\onlinecite{Fradkin1,Fradkin2}, and Sec.~IV.~of 
Ref.~\onlinecite{DasSarma-graphenereview}.

The zero-temperature thermal equilibrium state of the electrons in this sample is described by the zero-temperature Fermi-Dirac distribution $f_0(\varepsilon_{\trm{c} \vec k})=\Theta(\varepsilon_\trm{F}-\varepsilon_{\trm{c}\vec k}) $.
The initial state we consider is depicted in Fig.~\ref{fig1}a.
The distribution function in the initial state (Fig.~\ref{fig1}a) can be formulated using valley-dependent Fermi energies $\varepsilon_{\trm{F}K}$ and $\varepsilon_{\trm{F}K'}$: 
\begin{subequations}
\label{eq:init}
\bean
f_{\vec K + \vec \kappa}\left(t=0\right) &=& \Theta(\varepsilon_{\trm{F}K} - \varepsilon_{\trm{c},\vec K+\vec \kappa}), \\
f_{\vec K'+ \vec \kappa}\left(t=0\right) &=& \Theta(\varepsilon_{\trm{F}K'} -\varepsilon_{\trm{c},\vec K'+\vec \kappa}).
\eean
\end{subequations}
In line with Fig.~\ref{fig1}a, we assume $\varepsilon_{\trm{F}K'}<\varepsilon_{\trm{F}} < \varepsilon_{\trm{F}K}$; due to particle conservation, only two of the three Fermi energies are independent. Furthermore, we consider an initial state where the occupation difference in the two valleys is small, in the sense that all the scattering rates of the states participating in the dynamics can be approximated by scattering rates at the Fermi energy 
$\varepsilon_\trm{F}$.

Our aim here is to describe the relaxation dynamics of the valley polarization density $n_{\trm v}(t)$, which is related to the distribution function via 
\bean
\label{eq:valleypoldens}
n_{\trm v}(t) = \frac{1}{A} \sum_{\vec \kappa} \left[ f_{\vec K+\vec \kappa}(t)
- f_{\vec K' + \vec \kappa}(t)\right],
\eean
where $A$ is the sample area. This relation suggests that in order to obtain $n_{\trm v}(t)$, it is not necessary to solve the original Boltzmann equation \eqref{eq:fBoltzmann}. Instead, formulating and solving a time-evolution equation for the distribution difference
\bean
\label{eq:diff}
f^\trm{(v)}_{\vec \kappa}  
=f_{\vec K+\vec \kappa}
- f_{\vec K' + \vec \kappa}
\eean
might be sufficient. This can be done if the conditions
\begin{subequations}
\label{eq:Wsym}
\bean
\label{eq:Wsym1}
W_{\vec K + \vec \kappa,\vec K+\vec \kappa'}
&=&
W_{\vec K' + \vec \kappa,\vec K'+\vec \kappa'}
\equiv W^{(KK)}_{\vec \kappa \vec \kappa'}
\\
\label{eq:Wsym2}
W_{\vec K + \vec \kappa,\vec K'+\vec \kappa'}
&=&
W_{\vec K' + \vec \kappa,\vec K+\vec \kappa'}
\equiv W^{(KK')}_{\vec \kappa \vec \kappa'}
\eean
\end{subequations}
are fulfilled. Eqs.~\eqref{eq:Wsym1} and \eqref{eq:Wsym2}
describe the intravalley and intervalley transition rates, respectively.
From now on, we rely on these conditions, and in Appendix \ref{app:boltzmann_cond} we argue that they are indeed approximately fulfilled under the small-Fermi-energy condition [Eq.~\eqref{eq:lincond}].

A straightforward calculation using Eqs.~\eqref{eq:fBoltzmann}, \eqref{eq:Wsym1} and \eqref{eq:Wsym2} shows that the time-evolution equation for $f^\trm{(v)}_{\vec \kappa}$ reads
\bean
\frac{\partial f^{(\trm{v})}_{\vec{\kappa}}}{\partial t}&=&\sum_{\vec{\kappa}'} W^{(KK)}_{\vec{\kappa}\vec{\kappa}'}\left(f^{(\trm{v})}_{\vec{\kappa}'}-f^{(\trm{v})}_{\vec{\kappa}}\right)\nonumber \\&-& 
\sum_{\vec{\kappa}'} W^{(KK')}_{\vec{\kappa}\vec{\kappa}'}\left(f^{(\trm{v})}_{\vec{\kappa}'}+f^{(\trm{v})}_{\vec{\kappa}}\right)
\label{eq:Boltzmann_valley}
\eean
Importantly, the initial distribution difference $f_{\vec \kappa}^\trm{(v)}(0)$, defined via Eqs.~\eqref{eq:init} and \eqref{eq:diff}, is approximately isotropic in $\vec \kappa$ in due to our small-Fermi-energy condition
[Eq.~\eqref{eq:lincond}]. 
Therefore the Boltzmann equation \eqref{eq:Boltzmann_valley}
is solved by the time-evolving distribution function
\bean
\label{eq:solution}
f^{(\trm{v})}_\vec{\kappa}\left(t\right)&=&f^{(\trm{v})}_\vec{\kappa}\left(0\right)\mrm{e}^{-\Gamma_\trm{v} t}, 
\eean
with 
\bean
\Gamma_\trm{v}&=&2\sum_{\vec{\kappa}'}W^{(KK')}_{\vec{\kappa}\vec{\kappa}'},
\label{eq:valleyrelaxtime1}
\eean
provided that the sum on the right hand side of Eq.~\eqref{eq:valleyrelaxtime1}
is independent of the direction of $\vec \kappa$. 
This latter condition is fulfilled in the case we consider, 
and this becomes apparent when evaluating the
integral in Eq.~\eqref{eq:Sintegral} below. 

Combining Eqs.~\eqref{eq:valleypoldens}, \eqref{eq:diff}, and \eqref{eq:solution}, 
we find that 
the valley polarization density also shows an exponential decay:
\bean
n_{\trm v}(t) = n_\trm{v}(0) \mrm{e}^{-\Gamma_\trm{v} t}.
\eean
Therefore, we call $\Gamma_\trm{v}$ the \emph{valley relaxation rate} and
$\tau_\trm{v}=\Gamma_\trm{v}^{-1}$ the \emph{valley relaxation time}. Note that according to Eq.~\eqref{eq:valleyrelaxtime1}, the valley relaxation rate is twice as large as the intervalley scattering rate $\sum_{\vec \kappa'} W_{\vec \kappa \vec \kappa'}^{(KK')}$.

\subsection{Dispersion, wave functions and scattering rates}
\label{subsec:wfn}

In the previous section, we 
established the relation between the
valley relaxation time $\tau_\trm{v}$ and 
the intervalley scattering rates $W_{\vec \kappa \vec \kappa'}^{( K  K')}$.
The latter can be obtained
 from Fermi's Golden Rule in Eq.~\eqref{eq:FGR}, if the 
Bloch-type electronic wave 
functions $\psi_{\trm{c}\vec k}(\vec r)$ 
and the dispersion relation $\varepsilon_{\trm{c} \vec k}$
 of the conduction band are known.
Here, we outline the model we use to evaluate these quantities. 

We use the standard tight-binding or LCAO (linear combination of atomic orbitals) model of the $\pi$ electrons of graphene, with 
a special feature that the atomic $p_z$ orbitals are represented by normalized 2D Gaussian-like wave functions,
\bean
\label{eq:atomicwfn}
\phi\left(\vec{r}\right)=\frac{1}{\sqrt{2\pi}a_\trm{eB}}\mrm{e}^{-\frac{r^2}{4a_\trm{eB}^2}},
\eean
where $ a_\trm{eB} $ is the characteristic length scale of the orbitals, and the label $\trm{eB}$ corresponds to `effective Bohr radius'. We use $a_\trm{eB}$ as a parameter and assume that it is approximately an order of magnitude smaller than the carbon-carbon distance. 
Following Ref.~\onlinecite{WallacePR}, 
we will disregard the overlap of atomic wave functions centered on different
atoms.
This is a reasonable approximation as long as 
$a_\textrm{eB} \leq 0.3 \, a_\textrm{CC}$,
as the latter inequality 
guarantees that the nearest-neighbor overlap integral is less than $0.25$.
(For a free-standing carbon atom, we estimate that the 
spatial extension of a three-dimensional $2p_z$ atomic orbital in the xy plane 
is approximately 0.28$\,a_\textrm{CC}$, see Appendix \ref{app:estimateaeb}.)

With the above simplifications, the electronic 
Bloch wave functions are expressed as
\bnen
\ket{n\vec{k}} = \frac{1}{\sqrt{N_\trm{c}}} \sum_{\vec{R}} \mrm{e}^{\mrm{i}\vec{k}\vec{R}}\left( a_{n\vec{k}} \ket{\phi_\vec{R}^\trm{A}} + b_{n\vec{k}} \ket{\phi_\vec{R}^\trm{B}}\right),
\label{eq:wavefunction}
\eden
where $N_\trm{c}$ is the number of the unit cells in the sample,
$\braket{\vec{r}|\phi_\vec{R}^\trm{A}}=\phi\left(\vec{r}-\vec{R}\right) $ and $ \braket{\vec{r}|\phi_\vec{R}^\trm{B}} =\phi\left(\vec{r}-\vec{R}-\vec{\tau}\right) $, $\trm{A}$ and $\trm{B}$ are the two sublattices, $\vec{R}$ is a lattice vector, $a_{n\vec{k}}$ and $b_{n\vec{k}}$ are the sublattice amplitudes that can be obtained from the tight-binding model\cite{WallacePR} as
\bean
\left(\begin{matrix} a_{n\vec{k}}\\ b_{n\vec{k}}\\ \end{matrix}\right)&=\dfrac{1}{\sqrt{2}}\left(\begin{matrix} \frac{n \mathfrak{f}\left(\vec{k}\right)}{\left|\mathfrak{f}\left(\vec{k}\right)\right|}\\ 1\\ \end{matrix}\right)
\label{eq:sublatamps}
\eean
with $\mathfrak{f}(\vec{k})=1+\mrm{e}^{-\mrm{i}\vec{k}\vec{a}_1}+\mrm{e}^{-\mrm{i}\vec{k}\vec{a}_2}$. 
Here, the band index $n\in (+1,-1) \equiv (\trm{c},\trm{v})$ refers to the conduction ($+1$ or $\trm{c}$) and valence ($-1$ or $\trm{v}$) bands. 
Note that the vector $\vec \tau$,  and the primitive lattice vectors $\vec a_1$ and $\vec a_2$, are defined in Appendix \ref{sec:conventions},
and the direct lattice, the reference frame and the reciprocal lattice
are shown in Fig.~\ref{fig8} of Appendix \ref{sec:conventions}.
The dispersion relation reads
\bnen
\varepsilon_{n\vec k}=n\varepsilon_{\vec k}=n\gamma_0|\mathfrak{f}(\vec k)|,
\label{eq:disp}
\eden
where $\gamma_0\approx 2.8\,\trm{eV}$ is the nearest neighbour hopping matrix element.

In the vicinity of the two Dirac points, $\mathfrak{f}(\vec k)$ can be linearized, yielding the approximate sublattice amplitudes
\begin{subequations}
\bean
\label{eq:coeff1}
\left(\begin{matrix} a^{(\trm{lin})}_{n\vec{K}+\vec{\kappa}}\\ b^{(\trm{lin})}_{n\vec{K}+\vec{\kappa}}\\ \end{matrix}\right)&=&\frac{1}{\sqrt{2}}\left(\begin{matrix} \vphantom{a^{(\trm{lin})}_{n\vec{K}+\vec{\kappa}}} n\mrm{e}^{-\mrm{i}\varphi}\\ \vphantom{a^{(\trm{lin})}_{n\vec{K}+\vec{\kappa}}}1\\ \end{matrix}\right) \\
\label{eq:coeff2}
\left(\begin{matrix} a^{(\trm{lin})}_{n\vec{K}'+\vec{\kappa}}\\ b^{(\trm{lin})}_{n\vec{K}'+\vec{\kappa}}\\ \end{matrix}\right)&=&\frac{1}{\sqrt{2}}\left(\begin{matrix} \vphantom{a^{(\trm{lin})}_{n\vec{K}+\vec{\kappa}}} -n\mrm{e}^{\mrm{i}\varphi}\\ \vphantom{a^{(\trm{lin})}_{n\vec{K}+\vec{\kappa}}}1\\ \end{matrix}\right),
\eean
\end{subequations}
where $\varphi=\angle\left(\vec{\kappa},\hat{\vec{x}}\right)$ is 
the polar angle of $\vec \kappa$, 
and the linearized dispersion relation  
\bnen
\varepsilon^{(\trm{lin})}_{n\vec{K}+\vec{\kappa}}=\varepsilon^{(\trm{lin})}_{n\vec{K}'+\vec{\kappa}}=n \hbar v_\trm{F} \kappa,
\label{eq:lindisp}
\eden
where $v_\trm{F}=3 \gamma_0 a_\trm{CC} /2\hbar \approx 9\times10^5\,\trm{m/s}$ is the Fermi velocity in graphene.

Using the 2D Gaussian-like atomic wave function \eqref{eq:atomicwfn} in our tight-binding model is a simplification. In principle, one could use more realistic models, e.g., the linear combination of three-dimensional hydrogen-type atomic orbitals, or Bloch wave functions obtained from a numerical density-functional calculation.
As we show below, our choice \eqref{eq:atomicwfn} has the advantage that it yields simple analytical expressions for the calculated quantities,
including the inverse of the dielectric matrix at wave vector $\vec K$.
Furthermore, although our simplified approach might not yield 
quantitatively accurate results, it is expected to reveal the qualitative role
of the atomic wave function in the intervalley scattering processes, 
and to be used as a benchmark for future numerical approaches.

\section{Results}

\subsection{Unscreened impurities}
\label{sec:unscreened}

Here, we consider the case when the screening of the impurity
Coulomb potential due to electron-electron interaction is 
disregarded. 
Using the scattering rate in Eq.~\eqref{eq:FGR} and the isotropic linear spectrum around the Dirac points in Eq.~\eqref{eq:lindisp}, we obtain a formula from Eq.~\eqref{eq:valleyrelaxtime1} for the valley relaxation rate
\bnen
\label{eq:valleyrelaxtime}
\Gamma_\trm{v}=\frac{2\varepsilon_\trm{F} A}{\hbar^3 v_\trm{F}^2}\int_0^{2\pi}{\frac{\mrm{d}\varphi'}{2\pi}\left.\overline{\left|V_{\trm{c},\vec{K}+\vec{\kappa},\trm{c},\vec{K}'+\vec{\kappa}'}\right|^2}\right|_{\kappa'=\kappa=\kappa_\trm{F}}},
\eden
where $\varphi'=\angle\left(\vec{\kappa}',\hat{\vec{x}}\right)$
is the polar angle of $\vec \kappa'$.

The scattering matrix element can be obtained
using  the electronic wave function in Eq.~\eqref{eq:wavefunction}.
By neglecting the contributions of integrals involving atomic wave functions
located at different sites, 
we find
\bean
\label{eq:Vkk}
V_{\trm{c},\vec{K}+\vec{\kappa},\trm{c},\vec{K}'+\vec{\kappa}'} &=&\frac{1}{A}\sum_\vec{G}{S_{\trm{c},\vec{K}+\vec{\kappa},\trm{c},\vec{K}'+\vec{\kappa}'}\left(\vec{K}+\Delta\vec{\kappa}+\vec{G}\right)} \nonumber \\  
&\times&{P\left(\vec{K}+\Delta\vec{\kappa}+\vec{G}\right)V^*\left(\vec{K}+\Delta\vec{\kappa}+\vec{G};d\right)}, \nonumber
\\  
\eean
where we use that $\vec{K}'-\vec{K}$ is equivalent with $\vec{K}$ on the reciprocal lattice. 
(Note that $ K = 4\pi/3\sqrt{3}a_\trm{CC} \approx 1.7\,\trm{\AA}^{-1}$, with $ a_\trm{CC}\approx 1.42\,\trm{\AA}$ being the carbon-carbon distance in the graphene lattice.)
Furthermore $\Delta\vec{\kappa}=\vec{\kappa}'-\vec{\kappa}$,  
$ P\left(\vec{q}\right)=\int\trm{d}^{2} \vec r \, \mrm{e}^{-\mrm{i}\vec{q}\vec{r}} \left|\phi\left(\vec{r}\right)\right|^{2}  = \mrm{e}^{-\frac 1 2 a_\trm{eB}^2 q^2}$ 
is the Fourier-transformed probability density of the 
atomic orbital (`form factor'), $ V\left(\vec{q};d\right)=\int\trm{d}^{2} \vec r \, \mrm{e}^{-\mrm{i}\vec{q}\vec{r}} V\left(\vec{r};d\right) $ 
is the Fourier-transformed impurity potential and 
\bean
\label{eq:structure}
S_{n\vec{k},n'\vec{k}'}\left(\vec{q}\right)=a_{n\vec{k}}^*a_{n'\vec{k}'}+b_{n\vec{k}}^*b_{n'\vec{k}'}\mrm{e}^{-\mrm{i}\vec{q}\vec{\tau}}
\eean 
is a factor from sublattice amplitudes (`structure factor'). 
We have also used that $P(\vec q)$ is real-valued,
which is implied by the cylindrical symmetry of the atomic
wave function, $\phi\left(\vec{r}\right)=\phi\left(r\right)$.

Because of our small-Fermi-energy assumption [Eq.~\eqref{eq:lincond}],
the condition $\kappa,\kappa' \ll K$ holds and
therefore $\Delta \kappa \ll K$.
That implies that Eq.~\eqref{eq:Vkk} can be well approximated 
by taking the limit $\Delta \vec \kappa \to 0$:
\bean
V_{\trm{c},\vec{K}+\vec{\kappa},\trm{c},\vec{K}'+\vec{\kappa}'}&\approx &
\frac{1}{A}\sum_\vec{G}{S^\trm{(lin)}_{\trm{c},\vec{K}+\vec{\kappa},\trm{c},\vec{K}'+\vec{\kappa}'}\left(\vec{K}+\vec{G}\right)}\nonumber \\
&\times&{P\left(\vec{K}+\vec{G}\right)V^*\left(\vec{K}+\vec{G};d\right)}.
\eean
Here, we kept $\vec \kappa$ and $\vec \kappa'$ in the lower
index of the structure factor; using the approximated sublattice amplitudes
Eqs.~\eqref{eq:coeff1} and \eqref{eq:coeff2} we
obtain
\bean
\label{eq:Slinexplicit}
S^\trm{(lin)}_{\trm{c},\vec{K}+\vec{\kappa},\trm{c},\vec{K}'+\vec{\kappa}'}(\vec K+\vec G)=\frac{
	\mrm{e}^{-\mrm{i} (\vec K + \vec G) \vec \tau}
	-\mrm{e}^{\mrm{i}(\varphi +\varphi')}
}
{2}.
\eean

The next step is to perform the disorder average in Eq.~\eqref{eq:valleyrelaxtime}. Note that $V(\vec q;d) = V_\trm{i}(\vec q;d)\sum_{j=1}^{N_{\trm{i}}}\mrm{e}^{-\mrm{i}\vec{q}\vec{r}_j}$, and the assumption of homogeneously and independently positioned impurities
implies 
\bnen
\overline{V^*\left(\vec{K}+\vec{G}\right)V\left(\vec{K}+\vec{G}'\right)}=N_\trm{i}V^2_\trm{i}\left(\vec{K}+\vec{G};d\right)\delta_{\vec{G}\vec{G}'}.
\label{eq:impavg}
\eden
Using Eq.~\eqref{eq:impavg}, we find
\bean
\overline{\left|V_{\trm{c},\vec{K}+\vec{\kappa},\trm{c},\vec{K}'+\vec{\kappa}'}\right|^2}&=&\frac{N_\trm{i}}{A^2}\sum_\vec{G}{\left|S^\trm{(lin)}_{\trm{c},\vec{K}+\vec{\kappa},\trm{c},\vec{K}'+\vec{\kappa}'}\left(\vec{K}+\vec{G}\right)\right|^2} \nonumber \\
&\times & {P^2\left(\vec{K}+\vec{G}\right) V^2_\trm{i}\left(\vec{K}+\vec{G};d\right)}.
\eean
Using Eq.~\eqref{eq:Slinexplicit} to evaluate 
the integral over the polar angle $\varphi'$ of $\vec{\kappa}'$,
yields
\bnen
\label{eq:Sintegral}
\int_0^{2\pi}{\frac{\mrm{d}\varphi'}{2\pi}\left|S^\trm{(lin)}_{\trm{c},\vec{K}+\vec{\kappa},\trm{c},\vec{K}'+\vec{\kappa}'}\left(\vec{K}+\vec{G}\right)\right|^2}=\frac{1}{2},
\eden
which implies that the valley relaxation rate reads
\bnen
\Gamma_\trm{v} =\dfrac{n_{\trm{i}}\varepsilon_\trm{F}}{\hbar^3 v^2_{\trm{F}}} \sum_{\vec{G}} P^2\left(\vec{K}+\vec{G}\right)V^2_{\trm{i}}\left(\vec{K}+\vec{G};d\right).
\label{eq:keyresult1}
\eden
Here, $ n_\trm{i}  = N_\trm{i}/A $ is the sheet density of the impurities. 
Equation \eqref{eq:keyresult1} is one of the three key results of this work.

Let us now discuss the parameter dependence of
the valley relaxation rate $\Gamma_\trm{v}$. 
First, we note that its dependence on $n_\trm{i}$ is linear as expected. 
Second, the valley relaxation rate $\Gamma_\trm{v}$ is 
proportional to the Fermi energy $\varepsilon_\trm{F}$.
This linear relation is a simple consequence of the fact that the number of final states in the $K'$ valley, that can be reached by scattering
from the $K$ valley, is proportional to the density of states 
$D(\varepsilon_\trm{F}) = \varepsilon_\trm{F} / 2 \pi \hbar^2 v_\trm{F}^2$ per spin per valley, which, in graphene, is proportional to $\varepsilon_\trm{F}$ itself. 
The property that the valley relaxation rate is proportional to the density of states at the Fermi energy is expected to be true for other multi-valley materials as well.
For example, in 2D monolayer transition-metal dicalcogenides, where the 
low-energy dispersion relation in the conduction band is parabolic, 
we expect a Fermi-energy-independent valley relaxation rate (for unscreened 
impurities).

Another characteristic feature of the result \eqref{eq:keyresult1} is the
$\vec G$ sum, where squared Fourier components
of the impurity potential are summed up with the weight function
$P^2$.
The appearance of the Fourier components 
$V_\trm{i}\left(\vec{K}+\vec{G}\right)$ is not surprising,
since in our model, the bulk wave functions $\ket{\trm{c}\vec k}$ are
not plane waves but Bloch-type wave functions, 
i.e.,  superpositions of plane waves with wave 
numbers $\vec k+ \vec G$.
Note also that the number of relevant terms in the $\vec G$ sum of
Eq.~\eqref{eq:keyresult1} is controlled by the effective Bohr radius
$a_\trm{eB}$. 
If the effective Bohr radius is increased, the momentum-space
weight function
$P^2(\vec q) = \mrm{e}^{-a^2_\trm{eB} q^2}$ becomes narrower, 
and the number of Fourier components that contribute significantly 
to the valley relaxation rate decreases. 
Consequently, the valley relaxation time grows with increasing
effective Bohr radius. 

\begin{figure}
	\includegraphics[width=1\columnwidth]{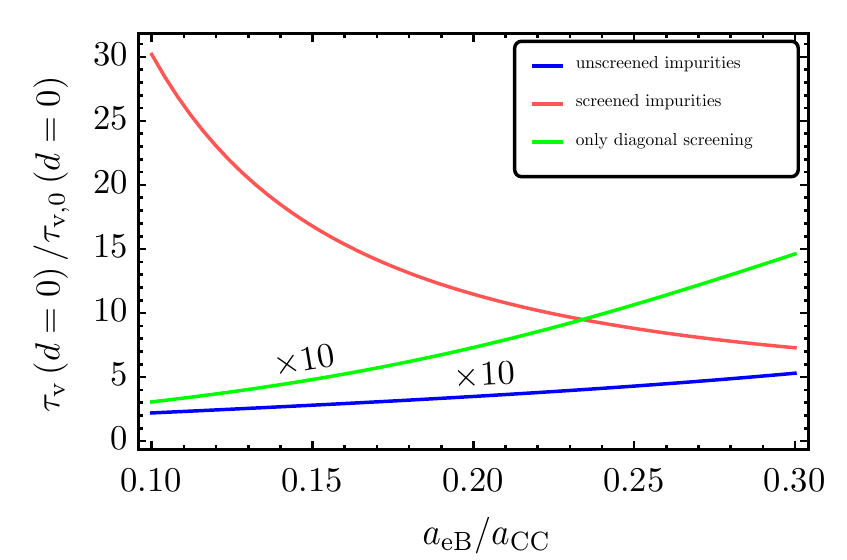}
	\caption{Valley relaxation time as a function of the
	effective Bohr radius. 
	The charged impurities are assumed to be located 
	in the graphene plane ($d=0$). 
	The cases of unscreened impurities (blue) and
	screened impurities (red) are shown.
	The green curve corresponds to a screened calculation where the off-diagonal
	matrix elements of the invrese dielectric matrix are disregarded (`diagonal 
	screening').
	The unit of the vertical axis is defined in Eq.~\eqref{eq:tauvpnull}. 
	The unit of the horizontal axis is the carbon-carbon distance $a_\trm{CC}$.}
	\label{fig3}
\end{figure} 

This trend is seen in Fig.~\ref{fig3}, where the 
valley relaxation time corresponding to the unscreened result  \eqref{eq:keyresult1}, for graphene-impurity distance $d=0$,
is shown as the blue line.
The result was obtained by numerically computing the sum in
Eq.~\eqref{eq:keyresult1}, which converges due to 
the Gaussian decay of $P^2(\vec k)$.
The unit on the vertical axis of Fig.~\ref{fig3} is defined 
as the $\vec G=\vec{0}$ term of the sum in Eq.~\eqref{eq:keyresult1}, i.e., 
\bnen
\tau_{\trm{v},0}(d) \equiv
\left[\dfrac{n_{\trm{i}}\varepsilon_\trm{F}}{\hbar^3 v^2_{\trm{F}}}V^2_{\trm{i}}\left(\vec{K};d\right)
\right]^{-1}
= \frac{\hbar^2 v_\trm{F}^2 K^2}{4\pi^2 e_0^2 n_\trm i \varepsilon_\trm{F}}
\mrm{e}^{2 K d},
\label{eq:tauvpnull}
\eden
where Eq.~\eqref{eq:unscreenedpotential} was used, furthermore, $\Gamma_{\trm{v},0}(d)=\tau_{\trm{v},0}^{-1}(d)$. For example, using the realistic parameter set $n_\trm{i}=10^{11}\,\trm{cm}^{-2}$, $\varepsilon_\trm{F}=0.1\,\trm{eV}$ and $d=0$, we find $\tau_\trm{v,0} \approx10\,\trm{ps}$.
Note that the corresponding transport lifetime for the same
parameter set is $\tau_\trm{tr} \approx5\,\trm{fs}$.

\begin{figure}
	\includegraphics[width=1\columnwidth]{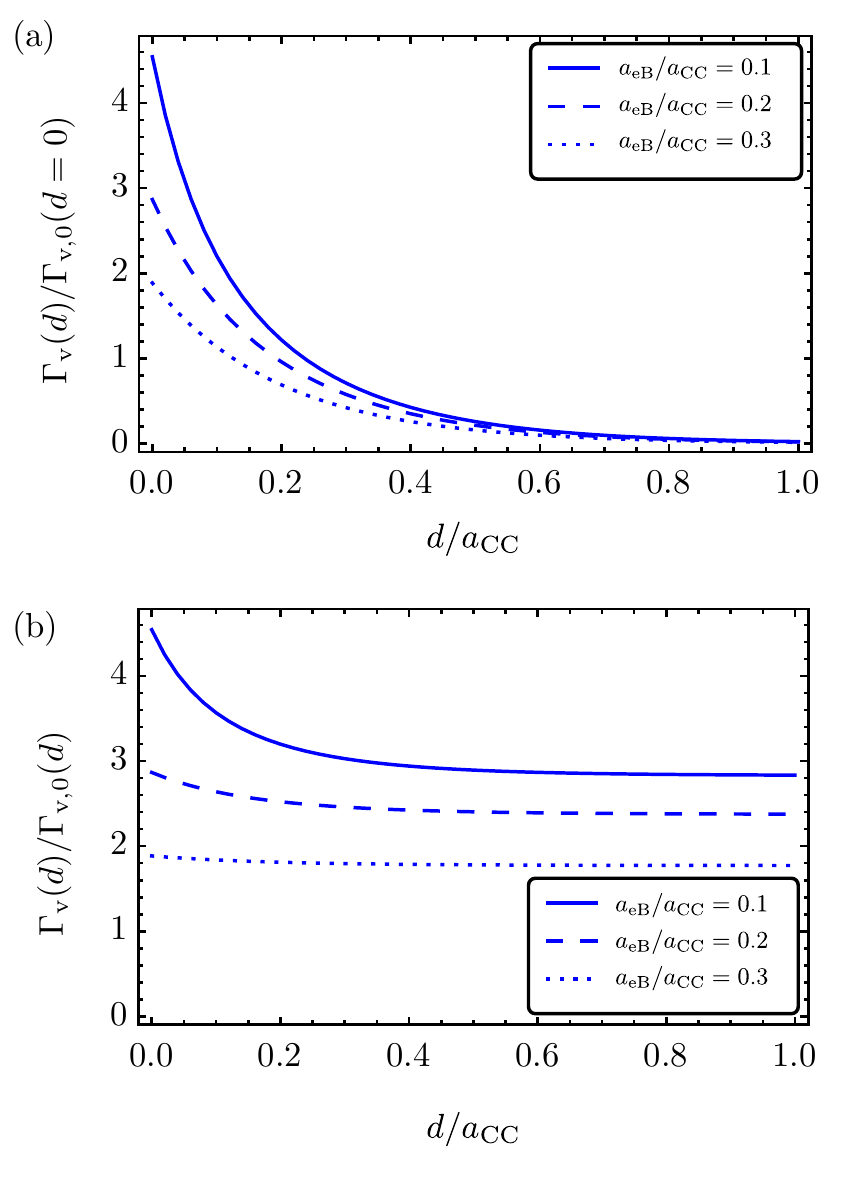}
	\caption{Valley relaxation rate as a function of graphene-impurity distance
	(unscreened impurities).
	(a) Approximately exponential decay of the valley relaxation rate with graphene-impurity distance. 
	Valley relaxation rate $\Gamma_\trm{v}(d)$ is shown for three different values of the effective Bohr radius.
	The unit of the vertical axis is $\Gamma_{\trm{v},0}$ at $d=0$ [see Eq.~\eqref{eq:tauvpnull}]. 
	(b) Deviations from exponential decay: same data as in (a), but
	here the ratio of the valley relaxation rate and the $d$-dependent
	 $\Gamma_{\trm{v},0}(d) \propto e^{-2Kd}$
	 [see Eq.~\eqref{eq:tauvpnull}] is plotted.
	 The plot indicates that for $d\lesssim a_\trm{CC}$, the decay
	 of the valley relaxation rate with $d$ does not follow
	 $\propto e^{-2Kd}$ exactly, and the deviation becomes less significant
	 as the effective Bohr radius is increased. 
	}
	\label{fig4}
\end{figure} 

For a given $d$, the value of $\tau_{\trm{v},0}(d)$ can be regarded as an order-of-magnitude estimate of the valley relaxation time. Equation \eqref{eq:tauvpnull} reveals that this estimate grows exponentially with graphene-impurity distance $d$, $\tau_{\trm{v},0} \propto \mrm{e}^{d/\ell}$, where the characteristic length scale of the growth is $\ell = 1/2K \approx 0.3\,\trm{\AA}$.
This approximately exponential dependence of the valley relaxation
time, and the above-estimated characteristic length 
scale, are illustrated in Fig.~\ref{fig4}a, where  $\Gamma_\trm{v}$ is shown as the function of the
graphene-impurity distance $d$.  
One implication of this approximately exponential behaviour is as follows. 
If the graphene layer is lying directly on a substrate, then charged impurities in the latter might be at a few-angstrom distance from the graphene plane, and can cause a relatively short valley relaxation time.  
For example, if $n_\trm{i}=10^{11}\,\trm{cm}^{-2}$, $\varepsilon_\trm{F}=0.1\,\trm{eV}$ and $d=3\,\trm{\AA}$, then $\tau_{\trm{v},0} \approx 200\,\trm{ns}$. However, if the graphene layer is suspended at a finite distance above the substrate, then the valley relaxation time improves exponentially with the distance; e.g., if $d = 3\,\trm{nm}$, then $\tau_{\trm{v},0}$ improves with a factor of $\approx 2 \times 10^4$.

A further thing to note about the $\vec G$ sum of Eq.~\eqref{eq:keyresult1}
is that the three terms corresponding to $\vec G =\vec{0}$,
$\vec G =-\vec b_1$ and $\vec G = \vec b_2$ are equal
(see Fig.~\ref{fig8} for the definitions of $\vec b_{1}$
and $\vec b_2$), 
because the corresponding three wave vectors $\vec K$,
$\vec K - \vec b_1$ and $\vec K+\vec b_2$ are
equal in length, and 
 $V_\trm{i}(\vec q)$ and $P(\vec q)$ are both
cylindrically symmetric. 
In fact, these three terms dominate the $\vec G$ sum in the case of
large graphene-impurity distance, $d\gtrsim a_\trm{CC}$,
as the remaining terms corresponding to longer wave vectors 
are suppressed due to the relation
$V_\trm{i}^2(\vec q;d) \propto \mrm{e}^{-2 q d}$, 
see Eq.~\eqref{eq:unscreenedpotential}.
As a consequence, $\tau_\trm{v}(d) \propto \tau_{\trm{v},0}(d)$ holds
for $d \gtrsim a_\trm{CC}$.
This relation is demonstrated in Fig.~\ref{fig4}b, where the
ratio $\tau_\trm{v}(d)/\tau_{\trm{v},0}(d)$ is indeed shown to saturate
for $d \gtrsim a_\trm{CC}$.
Figure \ref{fig4}b also shows that,
in the case of $d\lesssim a_\trm{CC}$, the proportionality  relation 
$\tau_\trm{v} \propto \tau_{\trm{v},0}$ breaks down,
especially for smaller values of the effective Bohr radius $a_\trm{eB}$.

\subsection{Screening within the random phase approximation}

So far, we have evaluated the valley relaxation time in the presence of 
unscreened impurities. 
Next, we improve on this result by taking into account
electron-electron interaction, which screens
the impurity-induced Coulomb potential.
In this subsection, we use the RPA approach to evaluate the dielectric matrix of graphene at
the intervalley wave number,
$\epsilon_{\vec G\vec G'}(\vec K)$;
then we use that to calculate the valley relaxation time in Sec.~\ref{sec:screened}. 

We note that most of the preceding theoretical works
studying dielectric screening by the $\pi$ electrons of 
graphene\cite{AndoJPSJ,HwangPRB-dielectric,WunschNJP} 
apply jellium-type descriptions, where the charge density corresponding 
to the Bloch-type wave functions is assumed to be homogeneous, 
the electronic dispersion relation is approximated by the Dirac cones,
the two valleys are treated independently, their contributions to screening
are simply added up, and screening is described by a dielectric
function $\epsilon(\vec q)$ instead of the dielectric matrix
we use below.  
This method might be appropriate as long as one is interested in the
screening of the long-wave-length Fourier components of the 
impurity potential. 
This is the case, e.g., when the conductivity
is calculated, since that is largely determined by intravalley scattering
processes.
However, here we consider intervalley scattering, 
which involves a momentum transfer comparable to the inverse 
of the atomic length scale. 
Therefore, it is required that the atomic-scale structure of the electronic
wave functions of the crystal, as well as the atomic-scale structure of the electrostatic potential (a.~k.~a.~\emph{local-field effects}), are taken into account, and that the wave-vector
summations are performed for the Brillouin zone.
We do these following the approach of Adler\cite{AdlerPR} and Wiser\cite{WiserPR}.
We note that the dielectric response of graphene has been described, with   local-field effects included, to characterise intervalley plasmons \cite{Tudorovskiy}, and the macroscopic static dielectric function\cite{Schilfgaarde}.

To characterise how the impurity Coulomb potential is screened
by the electrons in graphene, we need to express the
total (or screened) potential $V_\trm{tot}$ that has two
contributions:
the external potential $V_\trm{ext}$ created by the impurities, and
the induced potential created by the rearranged electrons. 
This relation is customarily expressed via the 
inverse dielectric function $\epsilon^{-1}(\vec r,\vec r')$ as
\bnen
\label{eq:realspaceinversedielectric}
V_\trm{tot}\left(\vec{r}\right)=\int{\mrm{d}^2 \vec r \,\epsilon^{-1}\left(\vec{r},\vec{r}'\right)V_\trm{ext}\left(\vec{r}'\right)}.
\eden
Graphene, being crystalline, has discrete translational invariance.
This implies that an external potential with wave vector $\vec q$ induces 
a total potential which is a superposition of Fourier components
with wave vectors $\vec q + \vec G$, 
where $\vec G$ is a reciprocal lattice vector. 
Hence the relation between the Fourier components of the
external and induced potential reads
\bnen
V_\trm{tot}\left(\vec{q}+\vec{G}\right)=\sum_{\vec{G}'}{\epsilon_{\vec{GG}'}^{-1}\left(\vec{q}\right)V_\trm{ext}\left(\vec{q}+\vec{G}'\right)},
\label{eq:matrixdielectricfunction}
\eden
where $\vec{q}$ is defined in the first Brillouin zone and $\vec{G}$, $\vec{G}'$ are reciprocal vectors.
The quantity $\epsilon^{-1}(\vec q)$ is called the 
\emph{inverse dielectric matrix}, 
its matrix elements are denoted by $\epsilon^{-1}_{\vec G \vec G'}(\vec q)$,
and these matrix elements are related to the Fourier components
of the inverse dielectric function via
\bean
\epsilon_{\vec{GG}'}^{-1}\left(\vec{q}\right)=\epsilon^{-1}\left(\vec{q}+\vec{G},-\vec{q}-\vec{G}'\right).
\label{eq:invdielmatnot} 
\eean
We emphasise that the inverse dielectric matrix 
is a matrix-valued function whose domain
is the first Brillouin zone. 
The \emph{dielectric matrix} $\epsilon(\vec q)$ is also a matrix-valued function on
the first Brillouin zone, fulfilling $[\epsilon(\vec q)]^{-1} = \epsilon^{-1}(\vec q)$
for all $\vec q$.

The  dielectric matrix is related to the
polarizability matrix $\Pi(\vec q)$ 
as (see Appendix \ref{app:RPA}):
\bnen
\epsilon_{\vec{G}\vec{G}'}\left(\vec{q}\right) = \delta_{\vec{G}\vec{G}'}-V_{\trm{C}}\left(\vec{q}+\vec{G}\right) \Pi_{\vec{G}\vec{G}'}\left(\vec{q}\right),
\label{eq:selfconsistantRPA}
\eden 
where $ \delta_{\vec{GG}'} $ is the Kronecker delta, and $ V_{\trm{C}}\left(\vec{q}+\vec{G}\right) = 2\pi e_0^2/|\vec{q}+\vec{G}| $ is the 2D Fourier transform of the Coulomb potential of the electron-electron interaction.

In the RPA, the polarizability matrix is approximated with that
of the noninteracting electron system.
The latter can be obtained from first-order static perturbation theory, and is expressed via the Adler-Wiser formula\cite{AdlerPR,WiserPR} 
\bean
\label{eq:adlerwiser}
\Pi_{\vec{G}\vec{G}'}\left(\vec{q}\right) &=& 
\frac{g_\trm{s}}{A}\sum_{nn'\vec{k}}
\frac
	{f\left(\varepsilon_{n\vec{k}}\right)-f\left(\varepsilon_{n'{\vec{k}+\vec{q}}}\right)}
	{\varepsilon_{n\vec{k}}-\varepsilon_{n'{\vec{k}+\vec{q}}}}
\\ \nonumber
&\times&
\braket{n\vec{k}|\mrm{e}^{-\mrm{i}\left(\vec{q}+\vec{G}\right)\vec{r}}|n'{\vec{k}+\vec{q}}} 
\braket{n'{\vec{k}+\vec{q}}|\mrm{e}^{\mrm{i}\left(\vec{q}+\vec{G}'\right)\vec{r}}|n\vec{k}},
\eean
where $g_\trm{s}=2$ accounts for the twofold spin degeneracy.
Note that Eq.~\eqref{eq:adlerwiser} is a generalisation of the 
Lindhard formula\cite{Lindhard}.
The latter expresses the polarizability of a homogeneous 
noninteracting system, whereas the former generalises that
to  the case of an inhomogeneous system with discrete translational
invariance. 
Furthermore, as we describe the regime of small Fermi energies, 
we will calculate and use the polarizability matrix of 
charge-neutral graphene (at zero temperature), which corresponds to 
$f(\varepsilon) = \Theta(-\varepsilon)$ in Eq.~\eqref{eq:adlerwiser}. 

Neglecting the contributions of integrals involving atomic wave functions
at different sites, the matrix elements in Eq.~\eqref{eq:adlerwiser}
can be simplified to
\bnen
\label{eq:matrixelement}
\braket{n\vec{k}|\mrm{e}^{-\mrm{i}\left(\vec{q}+\vec{G}\right)\vec{r}}|n'{\vec{k}+\vec{q}}}=P\left(\vec{q}+\vec{G}\right)S_{n\vec{k},n'\vec{k}+\vec{q}}\left(\vec{q}+\vec{G}\right),
\eden
where $P\left(\vec{q}+\vec{G}\right)$ and $S_{n\vec{k},n'\vec{k}'}\left(\vec{q}+\vec{G}\right)$ are defined above.

Recall that our goal is to use the inverse dielectric
matrix for calculating the valley relaxation rate.
To this end, we need to know the Fourier components
of the total potential in the vicinity of the wave vectors
$\vec K + \vec G$ only. 
Therefore, we need to evaluate the inverse dielectric matrix
in the vicinity of $\vec K$. 
Assuming that the inverse dielectric matrix is a smooth function,
we will calculate $\epsilon^{-1}(\vec K)$ explicitly and 
use the approximation
 $\epsilon^{-1}(\vec K+\vec{\Delta \kappa})\approx  \epsilon^{-1}(\vec K)$
 whenever $\vec K + \vec{\Delta \kappa}$ is close to $\vec K$ or an equivalent 
 wave vector.
Note that by using the notation $\epsilon^{-1}(\vec K)$ above, we have
implicitly defined $\vec K$ to be part of the first Brillouin zone.
Furthermore, we will call 
$\epsilon(\vec K)$,
$\epsilon^{-1}(\vec K)$,
and
$\Pi(\vec K)$
the \emph{intervalley dielectric matrix},
the \emph{inverse intervalley dielectric matrix}, 
and
the \emph{intervalley polarizability matrix}, respectively.

The intervalley polarizability matrix $\Pi(\vec K)$ can be expressed by invoking 
Eqs.~\eqref{eq:matrixelement}, \eqref{eq:adlerwiser}, \eqref{eq:structure}, and
\eqref{eq:sublatamps}, respectively. 
The result is 
\bnen
\Pi_{\vec{G}\vec{G}'}\left(\vec{K}\right) = \Pi_{\vec{K}}P\left(\vec{K}+\vec{G}\right)P\left(\vec{K}+\vec{G}'\right) \left[1+\mrm{e}^{\mrm{i}\left(\vec{G}'-\vec{G}\right)\vec{\tau}}\right],
\label{eq:chires}
\eden
where we defined
\bnen
\label{eq:chiK}
\Pi_{\vec{K}} = \frac{g_\trm{s}}{4A}\sum_{nn'\vec{k}}\frac{f\left(\varepsilon_{n\vec{k}}\right)-f\left(\varepsilon_{n'\vec{k}+\vec{K}}\right)}{\varepsilon_{n\vec{k}}-\varepsilon_{n'\vec{k}+\vec{K}}}.
\eden
Note that the quantity $\mathfrak{f}$ is absent from the result
\eqref{eq:chires}, even though $\mathfrak{f}$ appears in 
Eq.~\eqref{eq:sublatamps} describing the sublattice amplitudes;
the reason is that the $\vec k$ sum [in Eq.~\eqref{eq:adlerwiser}]
of the terms containing $\mathfrak{f}$ vanishes.

The quantity $\Pi_\vec K$ is evaluated 
assuming zero temperature $T=0$ and charge neutrality 
$\varepsilon_\trm{F}=0$ implying
$f(\varepsilon) = \Theta(-\varepsilon)$.
I.e., the conduction band is empty and the valence band is fully occupied, 
hence only the interband ($n \neq n'$) terms contribute to the
$n$, $n'$ sum in Eq.~\eqref{eq:chiK}:
\bnen
\Pi_{\vec{K}} = -\frac{g_\trm{s}}{2A}\sum_{\vec{k}}\frac{1}{\varepsilon_{\vec{k}}+\varepsilon_{\vec{k}+\vec{K}}},
\eden
where $\varepsilon_\vec{k}$ was defined in Eq.~\eqref{eq:disp}. 
To evaluate the summation over $\vec{k}$, we first
convert it to a momentum-space integral.
Despite the $\sim 1/k$ divergence of the integrand,
the integral is well defined due to its 2D nature. 
Numerical integration yields 
$\Pi_{\vec{K}}\approx -0.143/\gamma_0 a_\trm{CC}^2 \approx -2.53\times 10^{-2}\,\trm{1/eV\AA}^2$.

The intervalley dielectric matrix $\epsilon(\vec K)$ 
can be obtained using Eq.~\eqref{eq:selfconsistantRPA}
and Eq.~\eqref{eq:chires}. 
To determine the screened potential, we need to find the inverse
intervalley dielectric matrix $\epsilon^{-1}(\vec K)$, 
fulfilling the relation $ \sum_{\vec{G}''}{\epsilon^{\phantom{-1}}_{\vec{G}\vec{G}''}\left(\vec{K}\right)\epsilon^{-1}_{\vec{G''}\vec{G}'}\left(\vec{K}\right)}=\delta_{\vec{G}\vec{G}'}$.
We find that the matrix elements of $\epsilon^{-1}(\vec K)$ 
are 
\bnen
\epsilon^{-1}_{\vec{G}\vec{G}'}\left(\vec{K}\right) = 
\delta_{\vec{G}\vec{G}'}
+
\frac
	{V_{\trm{C}}\left(\vec{K} +\vec{G}\right) \Pi_{\vec G \vec G'}(\vec K)}
	{\epsilon_{\vec K}},
\label{eq:inversedielectric}
\eden
where we introduced the dimensionless quantity
\bnen
\label{eq:effectiveintervalley}
\epsilon_{\vec{K}}=1-\Pi_{\vec{K}} \sum_{\vec{G}}P^2\left(\vec{K}+\vec{G}\right) V_{\trm{C}}\left(\vec{K}+\vec{G}\right).
\eden
The analytical formula \eqref{eq:inversedielectric} for the inverse intervalley dielectric matrix is the second one of the three key results of this work. 

Let us close this subsection by discussing the qualitative features
of the results. 

First, focus on the intervalley polarizability matrix  $\Pi(\vec K)$ given
in Eq.~\eqref{eq:chires}.
(1) The characteristic scale of its
matrix elements is given by the quantity $\Pi_{\vec K}$, 
which is independent of the effective Bohr radius $a_\trm{eB}$. 
(2) The diagonal elements of $\Pi(\vec K)$ are real;
however,  the off-diagonal elements are complex in general, because of the
factor in the square brackets.
(3) The dependence of $\Pi_{\vec G \vec G'}(\vec K)$ on 
$a_\trm{eB}$ can be interpreted as follows. 
First, recall that 
the weight function $P(\vec q)$
is a 2D Gaussian function with a characteristic momentum-space width
of $1/a_\trm{eB}$. 
Therefore, Eq.~\eqref{eq:selfconsistantRPA}
testifies that an external potential of wave vector $\vec K + \vec G'$
can effectively 
create an induced electron density of wave vector $\vec K  + \vec G$,
if and only if both wave vectors are below $1/a_\trm{eB}$. 
If either $\vec K + \vec G$ or $\vec K + \vec G'$ are longer than $1/a_\trm{eB}$,
then either $P(\vec K+ \vec G)$ or
$P(\vec K  + \vec G')$ suppresses the matrix element 
$\Pi_{\vec G \vec G'}(\vec K)$. 
This makes sense: On the one hand, if  the wave vector $\vec K + \vec G'$
characterising the external potential is much longer then $1/a_\trm{eB}$, 
then the corresponding 
wave length is much shorter than $a_\trm{eB}$, hence the effect of the potential
on the electronic wave functions `averages out' and is therefore small indeed. 
On the other hand, the induced electron density is composed of atomic
orbitals, hence it cannot accommodate spatial variations with
smaller wave length than $a_\trm{eB}$.
Therefore its Fourier spectrum is constrained to the wave vectors shorter than
$1/a_\trm{eB}$, explaining 
the suppression factor $P(\vec K + \vec G)$ in Eq.~\eqref{eq:selfconsistantRPA}.

\begin{figure}[h]
	\includegraphics[width=1\columnwidth]{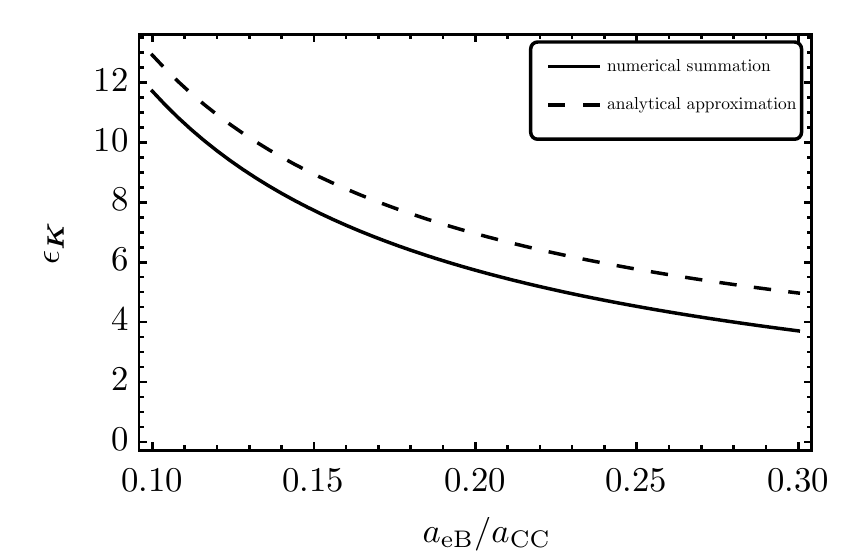}
	\caption{Effective intervalley dielectric constant as a function of the 
	effective Bohr radius.
	The unit of the horizontal axis is the carbon-carbon distance $a_\trm{CC}$.
	The results of numerical summation and analytical approximation are shown.}
	\label{fig5}
\end{figure}

Second, we consider the quantity $\epsilon_{\vec K}$ relevant
for the inverse intervalley dielectric matrix. 
By approximating the sum in Eq.~\eqref{eq:effectiveintervalley} to
an integral, we find 
\bnen
\epsilon_{\vec{K}}\approx 1-\Pi_{\vec{K}} \int{\frac{\mrm{d}^2 \vec k}{|\vec{b}_1\times\vec{b}_2|}P^2\left(\vec{k}\right) V_{\trm{C}}\left(\vec{k}\right)}\approx1+\frac{1.19}{a_\trm{eB}/a_\trm{CC}},
\label{eq:integral}
\eden
where $|\vec{b}_1\times\vec{b}_2|$ is the momentum-space 
area of the Brillouin zone. 
In Fig.~\ref{fig5}, we show $\epsilon_\vec{K}$ as a function the effective Bohr radius, as obtained via numerical evaluation of 
Eq.~\eqref{eq:effectiveintervalley} (solid line)
and via the integral approximation in Eq.~\eqref{eq:integral} (dashed line);
the two results show a reasonable qualitative agreement. 

\subsection{Screened impurity}
\label{sec:screened}

Here, we use our result \eqref{eq:inversedielectric}
for the inverse dielectric matrix to 
calculate the valley relaxation rate corresponding to screened 
charged impurities.
I.e., the disordered potential $V(\vec r;d)$ appearing in the
scattering rate \eqref{eq:FGR} is assumed to take the form 
[cf. Eq.~\eqref{eq:realspaceinversedielectric}]
\bean
V(\vec r;d) = \int d^2 \vec r'
\epsilon^{-1}(\vec r,\vec r')
\sum_{j=1}^{N_\trm{i}} V_\trm{i} (\vec r'- \vec r_j; d).
\eean
Repeating the calculation presented in Sec.~\ref{sec:unscreened} with this disorder potential, we find
\bean
\label{eq:tauvcalc}
\Gamma_\trm{v} &=&\dfrac{n_{\trm{i}}\varepsilon_\trm{F}}{\hbar^3 v^2_{\trm{F}}} \sum_{\vec{G}\vec{G}'\vec{G}''}\frac{1+\mrm{e}^{\mrm{i}\left(\vec{G}'-\vec{G}\right)\vec{\tau}}}{2}P\left(\vec{K}+\vec{G}\right)P\left(\vec{K}+\vec{G}'\right) \nonumber \\
&&\times \left[\epsilon^{-1}_{\vec{G}\vec{G}''}\left(\vec{K}\right)\right]^*  \epsilon^{-1}_{\vec{G}'\vec{G}''}\left(\vec{K}\right)V^2_{\trm{i}}\left(\vec{K}+\vec{G}'';d\right).
\eean
Substituting the formula of the intervalley dielectric matrix from Eq.~\eqref{eq:inversedielectric}, we obtain the following, remarkably simple analytical formula for the valley relaxation rate:
\bnen
\label{eq:keyresult3}
\Gamma_\trm{v} =\dfrac{n_{\trm{i}}\varepsilon_\trm{F}}{\hbar^3 v^2_{\trm{F}}} \sum_{\vec{G}} P^2\left(\vec{K}+\vec{G}\right)\left[\dfrac{V_{\trm{i}}\left(\vec{K}+\vec{G};d\right)} {\epsilon_{\vec{K}}}\right]^2.
\eden
Equation \eqref{eq:keyresult3} is the last one of the three key results
of this work. 

Note that the screened result \eqref{eq:keyresult3} 
can be obtained from the unscreened result \eqref{eq:keyresult1}
by substituting 
$V_\trm{i}\left(\vec{K}+\vec{G};d\right)/\epsilon_{\vec{K}}$
for 
$V_\trm{i}\left(\vec{K}+\vec{G};d\right)$.
On the one hand, it is remarkable that 
the matrix character 
of the inverse dielectric matrix in Eq.~\eqref{eq:inversedielectric} does not
appear explicitly in Eq.~\eqref{eq:keyresult3}; instead, 
the effect of screening on the valley relaxation rate is 
described by a single scalar $\epsilon_\vec{K}$ in 
Eq.~\eqref{eq:keyresult3}.
On the other hand, we 
emphasise that the role played by the quantity $\epsilon_{\vec{K}}$ in
the intervalley scattering rate is analogous to the role played by 
the dielectric constant in the screening of a long-wave-length
potential in a dielectric material.
Accordingly, $\epsilon_{\vec{K}}$ can be called the
\emph{effective intervalley dielectric constant}.
Importantly, the derivations of 
Eq.~\eqref{eq:keyresult1} and Eq.~\eqref{eq:keyresult3}
rely on our specific model for the electronic wave function 
(LCAO wave functions built from 2D Gaussian atomic wave functions) and
the specific type of disorder (random, uncorrelated Coulomb impurities); 
therefore, it is possible that the notion of the effective intervalley
dielectric constant is restricted to the present model only.

In Fig.~\ref{fig3}, the red curve shows the valley relaxation time
for screened Coulomb impurities, according to Eq.~\eqref{eq:keyresult3},
as a function of the effective Bohr radius $a_\trm{eB}$, 
for a graphene-impurity distance $d=0$. 
The $\vec G$ sum of Eq.~\eqref{eq:keyresult3} was evaluated numerically. 
The valley relaxation time is much longer in the screened case (red) than
in the unscreened case (blue): apparently, screening is effective in 
weakening the Coulomb potential of the impurities even at the
intervalley wave vector $\vec K$, and therefore
significantly prolongs the valley relaxation time.

Figure \ref{fig3} also shows that for screened impurities,
$\tau_\trm{v}$ is  decreasing with increasing effective Bohr radius, 
whereas for  unscreened impurities
$\tau_\trm{v}$ shows an opposite trend. 
As the only difference between the corresponding results
\eqref{eq:keyresult1} and \eqref{eq:keyresult3} is the appearance
of $\epsilon_{\vec K}$ in the latter, the different trends are explained by 
the relatively fast decay of $\epsilon_{\vec K}$ with increasing 
effective Bohr radius. 

At this point, the relative importance of the diagonal and off-diagonal matrix elements of the inverse intervalley dielectric matrix for the screened result (red) of Fig.~\ref{fig3} is not known. Using the hypothesis that only the diagonal matrix elements of $\epsilon^{-1}(\vec K)$ are important, we calculate the valley relaxation rate \eqref{eq:tauvcalc} of a screened impurity with an `artificial', diagonal inverse intervalley dielectric matrix, whose diagonal (off-diagonal) elements are given by $\epsilon^{-1}_{\vec G \vec G}(\vec K)$ of Eq.~\eqref{eq:keyresult3} (are zero). The obtained valley relaxation rate is shown in Fig.~\ref{fig3} as the green curve. The apparent qualitative difference between the screened result (red) and the diagonally screened result (green) reveals that the off-diagonal matrix elements of the inverse intervalley dielectric matrix do play an important role in the screening of the charged impurities.

\begin{figure}
	\includegraphics[width=1\columnwidth]{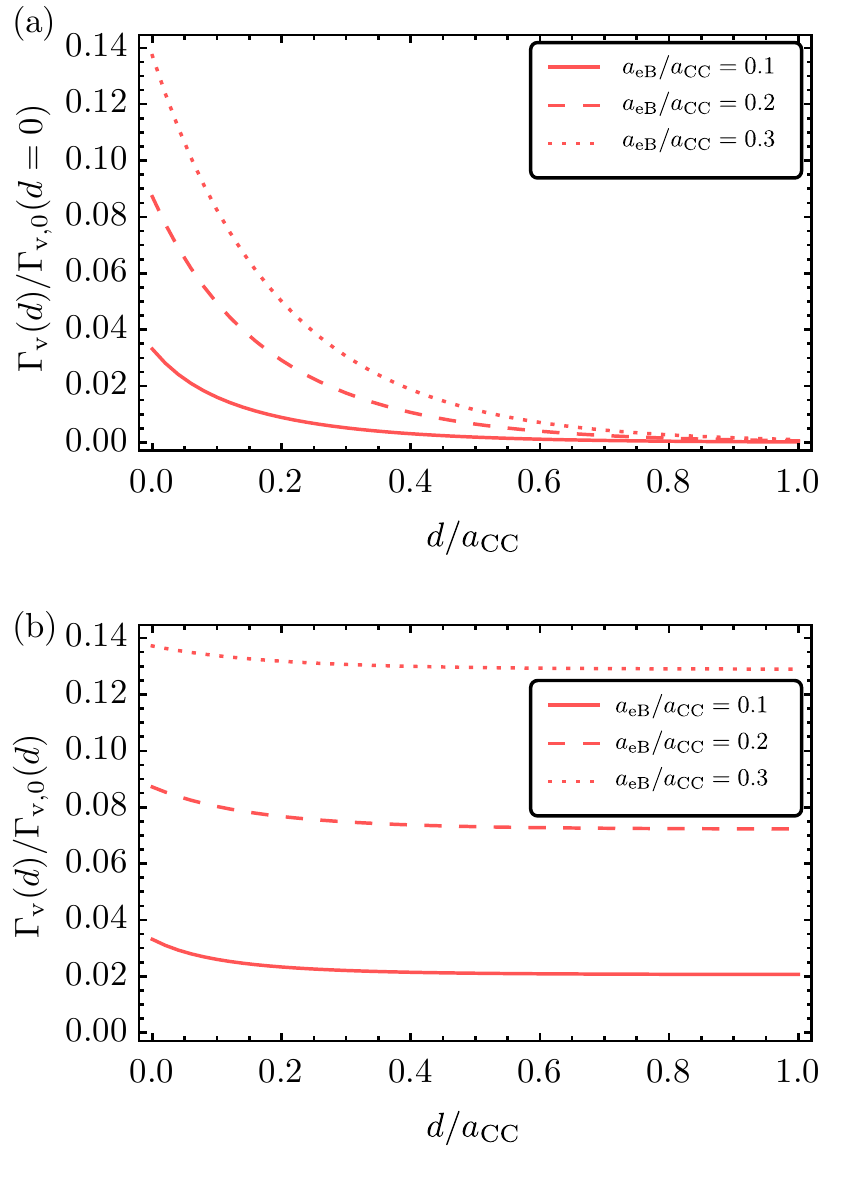}
	\caption{Valley relaxation rate as a function of graphene-impurity distance
	(screened impurities).
	See caption of Fig.~\ref{fig4} for more details.}
	\label{fig6}
\end{figure}

Finally, Fig.~\ref{fig6} shows the valley relaxation rate due to screened impurities as a function of the graphene-impurity distance, for three different values of the effective Bohr radius. The behavior is very similar to the unscreened case, see Fig.~\ref{fig4} and
the corresponding discussion in Sec. \ref{sec:unscreened}.

\section{Discussion and conclusions}

(1) In this work, we described the effects of screening  
using a model for the $\pi$ electrons of graphene, and we applied
the linear-response framework of the RPA.
It is a relevant, and, to our knowledge,  
open question how strongly
the other (e.g., $\sigma$) bands influence the dielectric matrix at 
wave vector $\vec K$.
The substrate might also contribute to 
the screening of the  short-wavelength Fourier components
of the Coulomb scatterers.
It is also an interesting future direction to describe how 
the effects beyond linear-response behaviour, e.g., 
bound-state formation around the Coulomb impurity\cite{Pereira,Shytov,Novikov}, 
influence the valley relaxation time. 
In a related recent theory work, intervalley scattering 
due to a combined long-range--short-range scatterer was studied, 
using a method where the effect of the long-range potential 
component was described nonperturbatively\cite{Braginsky}.

(2) 
In order to simplify calculations and to reveal the role  of the atomic structure of the electronic wave function in the valley relaxation process, we have used 2D Gaussian wave functions [Eq.~\eqref{eq:atomicwfn}] as building blocks of our model. 
This choice provided two advantages: (i) The 2D character of these atomic wave functions allows for a 2D description of screening effects. (ii) The Gaussian character allows one to derive simple analytical results. We wish to emphasize that advantage (i) is rather substantial. Without (i), the description of screening would become much more involved: graphene has discrete translation invariance within its plane, but has no  translational invariance in the out-of-plane direction. Hence, if 3D wave functions are used, then a `hybrid' theory should be developed for screening, which would then incorporate a special dielectric linear response function $\epsilon(\vec{q}+\vec G,-\vec{q}-\vec G',q_z,q'_z)$ [cf.~Eq.~\eqref{eq:invdielmatnot}], where $\vec G$ and $\vec G'$ are 2D reciprocal lattice vectors, $\vec q$ is a wave vector from the 2D Brillouin zone, and $q_z$ and $q'_z$ are out-of-plane wave numbers. Developing such a hybrid theory would be a welcome development, but we do not attempt that in the present work. Advantage (ii) is not that substantial. In fact, all our results expressed with the form factor $P(\vec q)$ hold for any other 2D atomic wave function as well, as long as the latter is cylindrically symmetric. 

(3) Here, we described valley relaxation due to 
scattering off long-range Coulomb impurities.
In general, the valley relaxation time is set by the interplay of a number of 
mechanisms (e.g., electron-phonon scattering, 
scattering off short-range impurities, etc.).

(4) Our calculation is restricted to zero temperature. 
At finite temperature, the valley relaxation rate is expected to change.
One possible reason for a temperature-dependent valley
relaxation rate is temperature-dependent screening: 
the temperature-dependent electronic distribution $f(\epsilon)$
appears in the formula for the polarizability matrix, see Eqs.~\eqref{eq:chires} and \eqref{eq:chiK}.
Another possible reason
causing temperature-dependent $\tau_\trm{v}$ is
the thermal population of phonons with large wave vectors,
that are capable to scatter electrons between the valleys upon
being absorbed.

(5) Here we assumed spatial homogeneity of the 
electronic distribution function $f$, 
and used the Boltzmann equation to describe the relaxation 
dynamics of a valley-polarized initial state.
A complementary task is to describe the valley
dynamics in a spatially inhomogeous structure, where, e.g., 
valley-polarized electrons are injected to a nanostructure
from a localised source\cite{Rycerz}.
We expect that our Boltzmann-equation-based approach 
can be used as a starting point in that case, 
to derive macroscopic transport equations 
describing valley diffusion, in analogy to the
spin-diffusion equations developed in spintronics\cite{Valet}.

\begin{figure}
	\includegraphics[width=1\columnwidth]{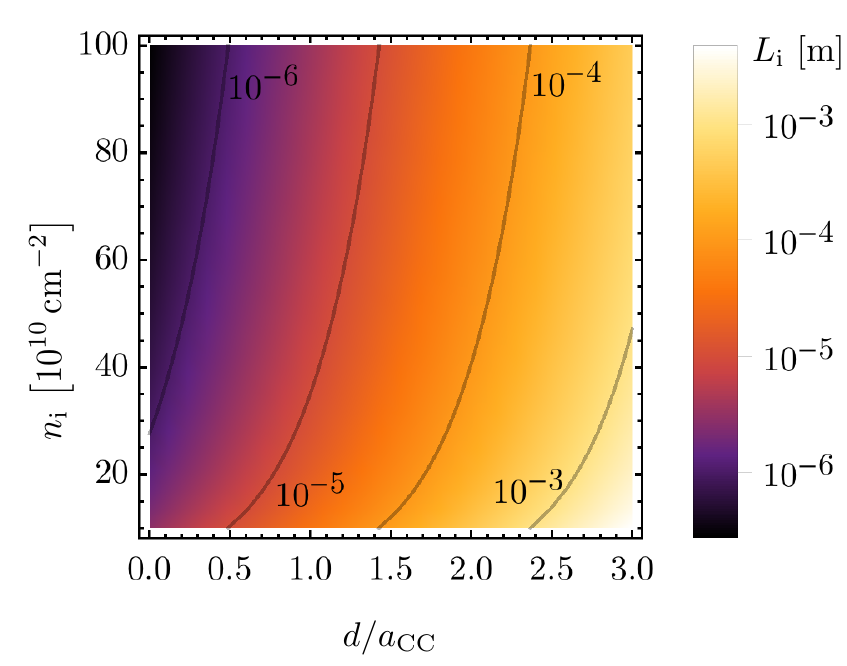}
	\caption{Valley-flip length as a function of the impurity sheet density and graphene-impurity distance. The effective Bohr radius is set to $a_\trm{eB}=0.2a_\trm{CC}$. The unit of horizontal axis is the carbon-carbon distance $a_\trm{CC}$.
		\label{fig7}	}

\end{figure}

(6) The robustness of the valley index against scattering processes can be characterised by the \emph{valley-flip length} or \emph{intervalley scattering length} $L_\textrm{i}$: this is the typical distance an electron can travel without having its valley index flipped. In our case of charged impurities, intravalley scattering happens more often than intervalley scattering, hence the motion of the electron between two valley flips is diffusive.
Our results for the valley relaxation time $\tau_\trm{v}$ allow us to estimate the dependence of the valley-flip length as a function of system parameters. 
The diffusion coefficient is estimated as $D=\tau_\trm{tr}v_\trm{F}^2/2$, where $\tau_\trm{tr}$ is the transport lifetime.
Then the valley-flip length is $L_\textrm{i} = \sqrt{D\tau_\trm{i}}$, where $\tau_\trm{i} = 2 \tau_\textrm{v}$ is the intervalley scattering time, see Eq.~\eqref{eq:valleyrelaxtime1}.
As $\tau_\textrm{tr} \propto \varepsilon_\textrm{F}/n_\textrm{i} $ (see Ref.~\onlinecite{AndoJPSJ}) and $\tau_v \propto 1/n_\textrm{i} \varepsilon_\textrm{F}$ [see Eqs.~\eqref{eq:keyresult1} and \eqref{eq:keyresult3}], the valley-flip length is independent of the Fermi energy $\varepsilon_\trm{F}$ and inversely proportional to the impurity sheet density $n_\trm{i}$.
In Fig.~\ref{fig7}, we show the valley-flip length $L_\trm{i}$ as a function of the impurity sheet density and the graphene-impurity distance $d$, with the the effective Bohr radius set to $a_\trm{eB}=0.2a_\trm{CC}$; to obtain this result, Eq.~(3.23) of Ref.~\onlinecite{AndoJPSJ} was used for $\tau_\trm{tr}$ and our result
\eqref{eq:keyresult3} was used for $\tau_\trm{v}$.
Note that besides impurities, the rough edges of a real graphene sample can also induce intervalley scattering\cite{GrafNanoLett2007}. The relative importance of edge-induced and Coulomb-impurity-induced intervalley scattering can be judged by comparing the sample size and the valley-flip length evaluated in Fig.~\ref{fig7}. For example, for a sample size of 10 microns and graphene-impurity distance of $d=a_\trm{CC}$, Fig.~\ref{fig7} suggests that Coulomb scattering gains importance over edge scattering if $n_\trm{i} \gtrsim 3 \times 10^{11}\,\trm{cm}^{-2}$.

(7) In a recent measurement of the valley Hall effect in graphene\cite{GorbachevScience2014}, the length scale characterising the spatial decay of the non-local resistance was found to be $\approx 1.0 \, \mu$m. This length scale can be identified\cite{AbaninPRB2009} as the valley-flip length $L_\textrm{i}$ we defined above. Charged impurities are thought to be present in the measured device (see section 6 of the Supplementary Material of Ref.~\onlinecite{GorbachevScience2014}), hence it is motivated to relate this measured length scale to our theoretical results. Assuming that the charged impurity sheet density in the sample is in the range shown in our Fig.~7, the following interpretations can be suggested: the measured length scale $L_\textrm{i} \approx 1\, \mu$m is set by (i) charged impurities that are very close to the graphene plane ($d < 0.5 a_\textrm{CC}$), or (ii) sources other than charged impurities, e.g., edges or short-range impurities. 

In conclusion, we have presented a model for valley relaxation 
due to randomly positioned charged impurities in graphene. 
We described the dependence of the valley relaxation rate
of an ensemble of valley-polarized electrons on the model parameters
(Fermi energy, impurity sheet density, graphene-impurity distance, spatial extension of the atomic $p_z$ wave functions).
The static screening of the charged impurities was described by,
as required for crystalline materials, the dielectric matrix, 
which we evaluated in the RPA. 
Our results highlight that a quantitatively accurate description of valley relaxation 
is more challenging than that of the electrical conductivity: 
the former requires that screening due to electron-electron interaction
is described in terms of the dielectric matrix, and
that the spatial variation of the electronic wave functions 
on the atomic length scale is precisely known.

\begin{acknowledgments}
We thank Cs.~T\H{o}ke, G.~Sz\'echenyi, D.~Visontai and P.~Nagy for useful discussions.
We acknowledge funding from the EU Marie Curie Career Integration Grant CIG-293834 (CarbonQubits), the OTKA Grant PD 100373, and the 
EU ERC Starting Grant CooPairEnt 258789. 
A.~P.~is supported by the J\'anos Bolyai Scholarship of the Hungarian Academy of Sciences.
\end{acknowledgments}

\appendix
\section{Conventions}
\label{sec:conventions}

\begin{figure*}
	\includegraphics[width=17.8cm]{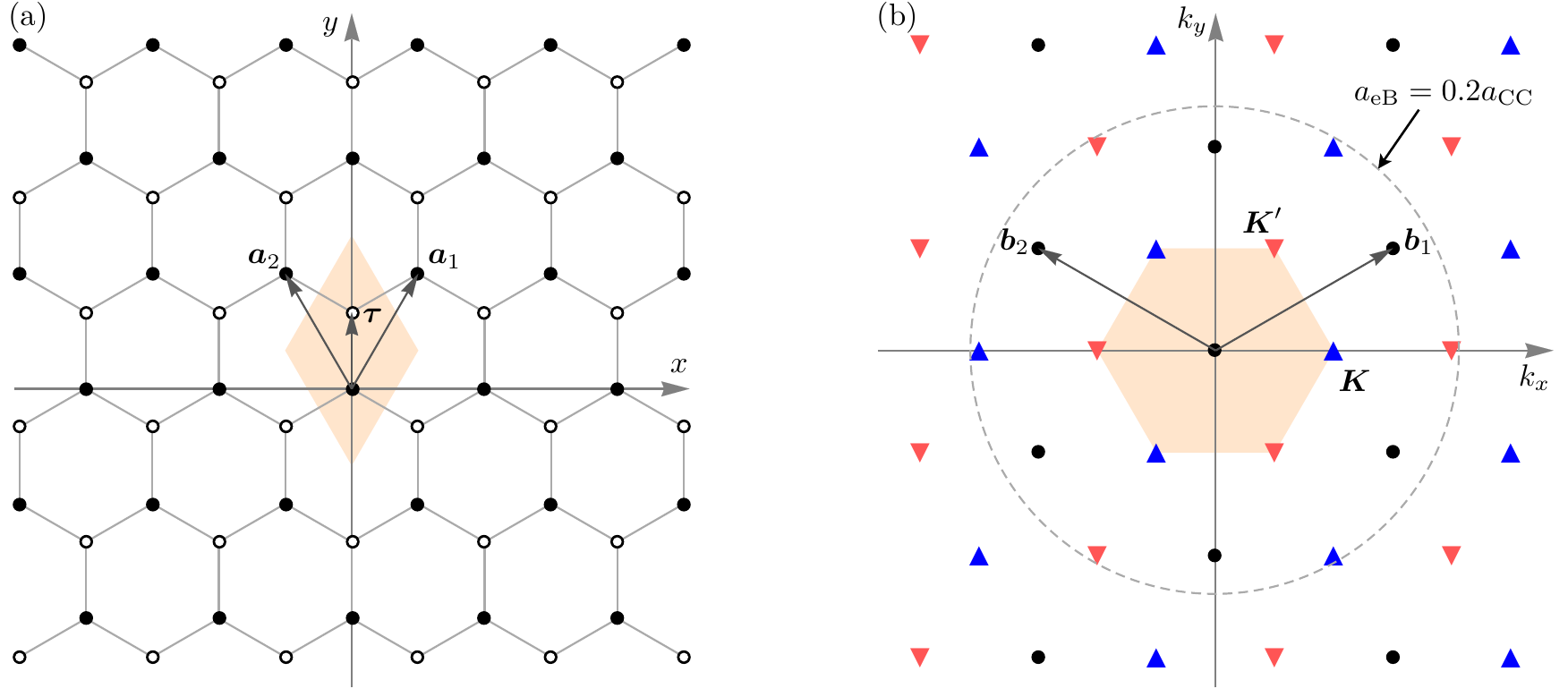}
	\caption{    \label{fig8}
    Direct and reciprocal lattices.
    (a) Direct lattice of graphene, showing the unit cell (shaded
    rhombus), the A (points)
    and B (circles) sublattices, the primitive lattice vectors
    $\vec a_1$ and $\vec a_2$, the vector $\vec \tau$ connecting
    the A and B sites within a unit cell,  and the reference frame.
    (b) Reciprocal lattice of graphene (black points), 
    showing the Brillouin zone (shaded hexagon), 
    the primitive reciprocal lattice vectors $\vec b_1$ and $\vec b_2$,
    the special points $\vec K$ and $\vec K'$ of the Brillouin zone,
    the shifted reciprocal lattices consisting of points $\vec K+ \vec G$
    and $\vec K' + \vec G$ (blue upward triangles and red downward
    triangles, respectively).
    The dashed circle shows the reciprocal-space extension
    of the Gaussian atomic electron density, i.e., 
    the set of momentum vectors where the 
    Fourier-transformed atomic electron density 
    is $\vec P(\vec k) = 1/2$, for an effective Bohr radius of 
    $a_\trm{eB} = 0.2 a_\trm{CC}$.     
    }
\end{figure*}

In this Appendix, the reference frame is specified and 
the vectors characterising the direct and reciprocal lattices are defined. 
The direct lattice is shown in Fig.~\ref{fig8}a. 
Atoms of the A and B sublattices are depicted as
black points and circles, respectively. 
The shaded rhombus shows the unit cell of the direct lattice. 
The primitive vectors of the direct lattice are
\begin{align}
\vec{a}_1 & =\frac{a_\trm{CC}}{2} \left(\begin{matrix} -\sqrt{3}\\ 3\\ \end{matrix}\right) &
\vec{a}_2 & =\frac{a_\trm{CC}}{2} \left(\begin{matrix} \sqrt{3}\\ 3\\ \end{matrix}\right) ,
\end{align}
and the vector connecting the A and B sites within a unit cell is
\begin{align}
\vec{\tau} & =a_\trm{CC} \left(\begin{matrix} 0\\ 1\\ \end{matrix}\right).
\end{align}
The primitive vectors of the reciprocal lattice are
\begin{align}
\vec{b}_1 & =\frac{2\pi}{3a_\trm{CC}} \left(\sqrt{3},1\right) &
\vec{b}_2 & =\frac{2\pi}{3a_\trm{CC}} \left(-\sqrt{3},1\right).
\end{align}
The Dirac points are
\begin{align}
\vec{K} & =\frac{4\pi}{3\sqrt{3}a_\trm{CC}} \left(1,0\right)&
\vec{K}' & =\frac{2\pi}{3\sqrt{3}a_\trm{CC}} \left(1,\sqrt{3}\right).
\end{align}

\section{Proof of Eqs.~\eqref{eq:Wsym1} and \eqref{eq:Wsym2}}
\label{app:boltzmann_cond}

First, we prove Eq.~\eqref{eq:Wsym1}, which corresponds to the
intravalley transition probability. 
The formula of the matrix element 
$V_{\trm{c},\vec K + \vec \kappa,\trm{c},\vec K+ \vec \kappa'}$ can be approximated by keeping only those terms that are proportional to $V(\Delta \vec \kappa)$, and dropping all other terms involving $V\left(\Delta\vec{\kappa}+\vec{G}\right)$  with $\vec{G}\neq0$ terms.
This is a reasonable approximation as our small-Fermi-energy condition implies $V(\Delta \vec \kappa) \sim 1/ \Delta \kappa \gg 1/G \sim V(\Delta \vec \kappa+ \vec G)$.
With this simplification, we find
\bnen
V_{\trm{c},\vec{K}+\vec{\kappa},\trm{c},\vec{K}+\vec{\kappa}'}\approx\frac{1}{A}S_{\trm{c},\vec{K}+\vec{\kappa},\trm{c},\vec{K}+\vec{\kappa}'}\left(\Delta\vec{\kappa}\right)P\left(\Delta\vec{\kappa}\right)  V^*\left(\Delta\vec{\kappa}\right),
\eden
 A similar formula holds for valley $K'$ with $S_{\trm{c},\vec{K}'+\vec{\kappa},\trm{c},\vec{K}'+\vec{\kappa}'}$. The transition probabilities are equal in the valleys if the structure factor has the property,
\bnen
\left|S_{\trm{c},\vec{K}+\vec{\kappa},\trm{c},\vec{K}+\vec{\kappa}'}\left(\Delta\vec{\kappa}\right)\right|^2=\left|S_{\trm{c},\vec{K}'+\vec{\kappa},\trm{c},\vec{K}'+\vec{\kappa}'}\left(\Delta\vec{\kappa}\right)\right|^2,
\eden
which can be proven by substituting Eqs.~\eqref{eq:coeff1} and \eqref{eq:coeff2} to the definition of structure factor \eqref{eq:structure} and approximating $\mrm{e}^{-\mrm{i}\Delta\vec{\kappa}\vec{\tau}}\approx 1$ due to $\Delta\kappa\rightarrow 0$. 
This argument can be reused for the case of screened impurities, 
assuming that long-wave-length screening is appropriately described
by the frequently used jellium-type RPA description\cite{AndoJPSJ,HwangPRB-dielectric,WunschNJP}.    

To prove Eq.~\eqref{eq:Wsym2}, which corresponds to the
intervalley transition rates, we rewrite Eq.~\eqref{eq:Wsym2} as
\bean
\label{eq:Wsym2app}
W_{\vec K + \vec \kappa,\vec K'+\vec \kappa'}
&=& W_{\vec K + \vec \kappa',\vec K'+\vec \kappa}=
W_{\vec K' + \vec \kappa,\vec K+\vec \kappa'},
\eean
where the second equality expresses detailed balance, which 
arises here as the transition rates are evaluated from Fermi's Golden Rule
and hence are invariant for the exchange of the initial and the final states. 
Thus, to confirm the first equality, it is enough to investigate the dependence of the intervalley transition rate $W_{\vec K+\vec \kappa,\vec K'+ \vec \kappa'}$ 
on the angles $\varphi$ and $\varphi'$ of $\vec \kappa$ and $\vec \kappa'$, respectively. 
The transition rate depends on $\varphi$ and $\varphi'$ only through $S^\trm{(lin)}_{\trm{c},\vec{K}+\vec{\kappa},\trm{c},\vec{K}'+\vec{\kappa}'}\left(\vec{K}+\vec{G}\right)$, which is invariant under the exchange of $\varphi$ and $\varphi'$polar angle of $\vec{\kappa}$ and $\vec{\kappa}'$ as seen from Eq.~\eqref{eq:Slinexplicit}. 
This proves the first equality in Eq.~\eqref{eq:Wsym2app}, and hence  \eqref{eq:Wsym2}, for both unscreened and screened impurities.

\section{Relation of the dielectric matrix and the polarizability matrix}
\label{app:RPA}

In this Appendix, we establish the relation \eqref{eq:selfconsistantRPA}
between the dielectric matrix and the polarizability matrix. 

First, recall that the real-space 
dielectric function $\epsilon(\vec r,\vec r')$
describes the relation between the 
external and total potentials [cf. Eq.~\eqref{eq:realspaceinversedielectric}],
\bean
\label{eq:realspacedielectric}
V_\trm{ext}(\vec r) = \int d^2\vec r' \epsilon(\vec r,\vec r') V_\trm{tot}(\vec r'),
\eean
whereas the real-space polarizability function $\Pi(\vec r,\vec r')$ 
describes the relation between 
the induced electron density and the total potential,
\bnen
\label{eq:realspacepolarizability}
n_\trm{ind}\left(\vec{r}\right)=\int{\mrm{d}^2 \vec r'\,\Pi\left(\vec{r},\vec{r}'\right)V_\trm{tot}\left(\vec{r}'\right)}.
\eden
For our purposes, $\vec r, \vec r'$ are 2D position vectors, since the
model we use to describe graphene's electrons is 2D.

Next, we apply Fourier transformation on the definitions 
\eqref{eq:realspacedielectric} and \eqref{eq:realspacepolarizability},
exploiting their invariance with respect to lattice translations:
e.g., 
$\Pi\left(\vec{r}+\vec{R},\vec{r}'+\vec{R}\right)=\Pi\left(\vec{r},\vec{r}'\right)$
for any lattice vector $\vec R$. 
Recall that we use the following definition for Fourier transformation
\bnen
f\left(\vec{k}\right)=\int{\mrm{d}^2r\,f\left(\vec{r}\right)\mrm{e}^{-\mrm{i}\vec{kr}}},
\eden
where $f(\vec r)$ is an arbitrary 2D position-dependent function.
We evaluate the Fourier transforms of Eqs.
\eqref{eq:realspacedielectric} and \eqref{eq:realspacepolarizability}
at an
arbitrary wave vector $\vec k = \vec q + \vec G$, where
$\vec q$ is within the  Brillouin zone and $\vec G$ is a reciprocal 
lattice vector; we obtain
\bean
V_\trm{ext}\left(\vec{q}+\vec{G}\right)
&=&
\sum_{\vec{G}'}{\epsilon_{\vec{GG}'}\left(\vec{q}\right)V_\trm{tot}\left(\vec{q}+
\vec{G}'\right)},
\label{eq:dielectricmatrix}
\\
n_\trm{ind}\left(\vec{q}+\vec{G}\right)
&=&
\sum_{\vec{G}'}{\Pi_{\vec{GG}'}\left(\vec{q}\right)V_\trm{tot}\left(\vec{q}+
\vec{G}'\right)},
\label{eq:indchitot}
\eean
Here, we used the fact that the invariance of the real-space response function
with respect to lattice translations implies, e.g., 
$\Pi(\vec q + \vec G, \vec q' + \vec G') \propto \delta_{\vec q,-\vec q'}$
and introduced the dielectric matrix $\epsilon(\vec q)$ 
and the polarizability matrix $\Pi(\vec q)$ 
as 
$\epsilon_{\vec{GG}'}\left(\vec{q}\right)=
\epsilon \left(\vec{q}+\vec{G},-\vec{q}-\vec{G}'\right)$ 
and
$\Pi_{\vec{GG}'}\left(\vec{q}\right)=
\Pi\left(\vec{q}+\vec{G},-\vec{q}-\vec{G}'\right)$,
respectively. 

Having these definitions at hand, the dielectric matrix and the 
polarizability matrix can be connected using Coulomb's law.
Coulomb's law implies that the induced potential, i.e., the potential
created by the induced electron density, reads
\bnen
V_\trm{ind}\left(\vec{r}\right)=\int{\mrm{d}^2 \vec r' \,V_\trm{C}\left(\vec{r}-\vec{r}'\right)n_\trm{ind}\left(\vec{r}'\right)},
\eden
where $V_\trm{C}\left(\vec{r}-\vec{r}'\right)=e_0^2/|\vec{r}-\vec{r}'|$ is the 2D Coulomb potential.
Applying Fourier transform with respect to $\vec{r}$ yields
\bnen
V_\trm{ind}\left(\vec{q}+\vec{G}\right)=V_\trm{C}\left(\vec{q}+\vec{G}\right)n_\trm{ind}\left(\vec{q}+\vec{G}\right),
\label{eq:Poissoneq}
\eden
where $V_\trm{C}\left(\vec{q}+\vec{G}\right)=2\pi e_0^2/|\vec{q}+\vec{G}|$ is the 2D Fourier transform of the Coulomb potential.
Inserting Eq.~\eqref{eq:Poissoneq} to the relation
\bnen
V_\trm{tot}\left(\vec{q}+\vec{G}\right)=V_\trm{ext}\left(\vec{q}+\vec{G}\right)+V_\trm{ind}\left(\vec{q}+\vec{G}\right),
\label{eq:totextind}
\eden
and eliminating $n_\trm{ind}$ via Eq.~\eqref{eq:indchitot}, 
one finds 
\bean
V_\trm{ext}\left(\vec{q}+\vec{G}\right)&=&\sum_{\vec{G}'}{\left[\delta_{\vec{GG'}}-V_\trm{C}\left(\vec{q}+\vec{G}\right)\Pi_{\vec{GG}'}\left(\vec{q}\right)\right]} \nonumber \\
&& \times V_\trm{tot}\left(\vec{q}+\vec{G}'\right),
\eean
proving Eq.~\eqref{eq:selfconsistantRPA}.

\section{Estimation of the effective Bohr radius}
\label{app:estimateaeb}

In the main text, the atomic $p_z$ orbital is represented by the 2D Gaussian-type wave function $\phi\left(\vec{r}\right)$, defined in Eq.~\eqref{eq:atomicwfn}. The 2D spatial extension of this wave function in the xy plane is characterized by the effective Bohr radius $a_\textrm{eB}$. Here, we estimate the 2D spatial extension of a three-dimensional 2$p_z$ orbital of a free-standing carbon atom, to provide an estimate for $a_\textrm{eB}$. 

To this end, we invoke those results of Ref.~\onlinecite{ClementiRaimondi} that correspond to a free-standing carbon atom. In Ref.~\onlinecite{ClementiRaimondi},  the three-dimensional single-electron orbitals are approximated by Slater-type orbitals; in particular, the 2$p_z$ orbital of a carbon atom is approximated by 
\bnen
\phi_{p_z}(\vec{r},z)=\frac{z}{\sqrt{32\pi(a_0/Z_\trm{eff})^5}}\mrm{e}^{-\frac{\sqrt{r^2+z^2}}{2a_0/Z_\trm{eff}}},
\label{eq:3Dorbit}
\eden
where $\vec{r}=(x,y)$, $r = \sqrt{x^2+y^2}$, 
$a_0=0.53\,\trm{\AA}$ is the Bohr radius, and 
$Z_\textrm{eff} = 3.14$ is the effective charge of the nucleus. 
(The latter is denoted in Ref.~\onlinecite{ClementiRaimondi} as $Z-\sigma$.)

We characterize the 2D spatial extension of 
$\phi_{p_z}$ in the xy plane via
\bean
\bra{\phi_{p_z}} r \ket{\phi_{p_z}} = \frac{15\pi}{16}a_0/Z_\trm{eff} \approx 0.50\,\textrm{\AA}
\eean
The same quantity for our 2D Gaussian-type wave function is
\bean
\bra{\phi} r \ket{\phi} = \sqrt{\frac{\pi}{2}}a_\trm{eB} .
\eean
From the requirement  
$\bra{\phi_{p_z}} r \ket{\phi_{p_z}} = \bra{\phi} r \ket{\phi}$,
we obtain the estimate 
\bnen
a_\trm{eB}=\frac{15}{8}\sqrt{\frac{\pi}{2}}a_0/Z_\trm{eff}\approx 0.40\,\trm{\AA}.
\eden
Expressing this result in units of the carbon-carbon distance $a_\textrm{CC}$, 
we find $a_\trm{eB} \approx 0.28\,a_\trm{CC}$.


\begin{thebibliography}{63}
\expandafter\ifx\csname natexlab\endcsname\relax\def\natexlab#1{#1}\fi
\expandafter\ifx\csname bibnamefont\endcsname\relax
  \def\bibnamefont#1{#1}\fi
\expandafter\ifx\csname bibfnamefont\endcsname\relax
  \def\bibfnamefont#1{#1}\fi
\expandafter\ifx\csname citenamefont\endcsname\relax
  \def\citenamefont#1{#1}\fi
\expandafter\ifx\csname url\endcsname\relax
  \def\url#1{\texttt{#1}}\fi
\expandafter\ifx\csname urlprefix\endcsname\relax\def\urlprefix{URL }\fi
\providecommand{\bibinfo}[2]{#2}
\providecommand{\eprint}[2][]{\url{#2}}

\bibitem[{\citenamefont{Rycerz et~al.}(2007)\citenamefont{Rycerz, Tworzydlo,
  and Beenakker}}]{Rycerz}
\bibinfo{author}{\bibfnamefont{A.}~\bibnamefont{Rycerz}},
  \bibinfo{author}{\bibfnamefont{J.}~\bibnamefont{Tworzydlo}},
  \bibnamefont{and} \bibinfo{author}{\bibfnamefont{C.~W.~J.}
  \bibnamefont{Beenakker}}, \bibinfo{journal}{Nat. Phys.}
  \textbf{\bibinfo{volume}{3}}, \bibinfo{pages}{172} (\bibinfo{year}{2007}).

\bibitem[{\citenamefont{Xiao et~al.}(2007)\citenamefont{Xiao, Yao, and
  Niu}}]{DiXiaoPRL2007}
\bibinfo{author}{\bibfnamefont{D.}~\bibnamefont{Xiao}},
  \bibinfo{author}{\bibfnamefont{W.}~\bibnamefont{Yao}}, \bibnamefont{and}
  \bibinfo{author}{\bibfnamefont{Q.}~\bibnamefont{Niu}},
  \bibinfo{journal}{Phys. Rev. Lett.} \textbf{\bibinfo{volume}{99}},
  \bibinfo{pages}{236809} (\bibinfo{year}{2007}).

\bibitem[{\citenamefont{Yao et~al.}(2008)\citenamefont{Yao, Xiao, and
  Niu}}]{WangYaoPRB2008}
\bibinfo{author}{\bibfnamefont{W.}~\bibnamefont{Yao}},
  \bibinfo{author}{\bibfnamefont{D.}~\bibnamefont{Xiao}}, \bibnamefont{and}
  \bibinfo{author}{\bibfnamefont{Q.}~\bibnamefont{Niu}},
  \bibinfo{journal}{Phys. Rev. B} \textbf{\bibinfo{volume}{77}},
  \bibinfo{pages}{235406} (\bibinfo{year}{2008}).

\bibitem[{\citenamefont{Gunlycke and White}(2011)}]{Gunlycke}
\bibinfo{author}{\bibfnamefont{D.}~\bibnamefont{Gunlycke}} \bibnamefont{and}
  \bibinfo{author}{\bibfnamefont{C.~T.} \bibnamefont{White}},
  \bibinfo{journal}{Phys. Rev. Lett.} \textbf{\bibinfo{volume}{106}},
  \bibinfo{pages}{136806} (\bibinfo{year}{2011}).

\bibitem[{\citenamefont{Tse et~al.}(2014)\citenamefont{Tse, Saxena, Smith, and
  Sinitsyn}}]{WangKongTse}
\bibinfo{author}{\bibfnamefont{W.-K.} \bibnamefont{Tse}},
  \bibinfo{author}{\bibfnamefont{A.}~\bibnamefont{Saxena}},
  \bibinfo{author}{\bibfnamefont{D.~L.} \bibnamefont{Smith}}, \bibnamefont{and}
  \bibinfo{author}{\bibfnamefont{N.~A.} \bibnamefont{Sinitsyn}},
  \bibinfo{journal}{Phys. Rev. Lett.} \textbf{\bibinfo{volume}{113}},
  \bibinfo{pages}{046602} (\bibinfo{year}{2014}).

\bibitem[{\citenamefont{Shan et~al.}(2015)\citenamefont{Shan, Zhou, and
  Xiao}}]{WenYuShanPRB2015}
\bibinfo{author}{\bibfnamefont{W.-Y.} \bibnamefont{Shan}},
  \bibinfo{author}{\bibfnamefont{J.}~\bibnamefont{Zhou}}, \bibnamefont{and}
  \bibinfo{author}{\bibfnamefont{D.}~\bibnamefont{Xiao}},
  \bibinfo{journal}{Phys. Rev. B} \textbf{\bibinfo{volume}{91}},
  \bibinfo{pages}{035402} (\bibinfo{year}{2015}).

\bibitem[{\citenamefont{Golub and Tarasenko}(2014)}]{Golub}
\bibinfo{author}{\bibfnamefont{L.~E.} \bibnamefont{Golub}} \bibnamefont{and}
  \bibinfo{author}{\bibfnamefont{S.~A.} \bibnamefont{Tarasenko}},
  \bibinfo{journal}{Phys. Rev. B} \textbf{\bibinfo{volume}{90}},
  \bibinfo{pages}{201402} (\bibinfo{year}{2014}).

\bibitem[{\citenamefont{Wehling et~al.}(2015)\citenamefont{Wehling, Huber,
  Lichtenstein, and Katsnelson}}]{WehlingPRB2015}
\bibinfo{author}{\bibfnamefont{T.~O.} \bibnamefont{Wehling}},
  \bibinfo{author}{\bibfnamefont{A.}~\bibnamefont{Huber}},
  \bibinfo{author}{\bibfnamefont{A.~I.} \bibnamefont{Lichtenstein}},
  \bibnamefont{and} \bibinfo{author}{\bibfnamefont{M.~I.}
  \bibnamefont{Katsnelson}}, \bibinfo{journal}{Phys. Rev. B}
  \textbf{\bibinfo{volume}{91}}, \bibinfo{pages}{041404}
  (\bibinfo{year}{2015}).

\bibitem[{\citenamefont{Zhu et~al.}(2012)\citenamefont{Zhu, Collaudin, Fauque,
  Kang, and Behnia}}]{Zhu}
\bibinfo{author}{\bibfnamefont{Z.}~\bibnamefont{Zhu}},
  \bibinfo{author}{\bibfnamefont{A.}~\bibnamefont{Collaudin}},
  \bibinfo{author}{\bibfnamefont{B.}~\bibnamefont{Fauque}},
  \bibinfo{author}{\bibfnamefont{W.}~\bibnamefont{Kang}}, \bibnamefont{and}
  \bibinfo{author}{\bibfnamefont{K.}~\bibnamefont{Behnia}},
  \bibinfo{journal}{Nat. Phys.} \textbf{\bibinfo{volume}{8}},
  \bibinfo{pages}{89} (\bibinfo{year}{2012}).

\bibitem[{\citenamefont{Cao et~al.}(2012)\citenamefont{Cao, Wang, Han, Ye, Zhu,
  Shi, Niu, Tan, Wang, Liu et~al.}}]{Cao}
\bibinfo{author}{\bibfnamefont{T.}~\bibnamefont{Cao}},
  \bibinfo{author}{\bibfnamefont{G.}~\bibnamefont{Wang}},
  \bibinfo{author}{\bibfnamefont{W.}~\bibnamefont{Han}},
  \bibinfo{author}{\bibfnamefont{H.}~\bibnamefont{Ye}},
  \bibinfo{author}{\bibfnamefont{C.}~\bibnamefont{Zhu}},
  \bibinfo{author}{\bibfnamefont{J.}~\bibnamefont{Shi}},
  \bibinfo{author}{\bibfnamefont{Q.}~\bibnamefont{Niu}},
  \bibinfo{author}{\bibfnamefont{P.}~\bibnamefont{Tan}},
  \bibinfo{author}{\bibfnamefont{E.}~\bibnamefont{Wang}},
  \bibinfo{author}{\bibfnamefont{B.}~\bibnamefont{Liu}}, \bibnamefont{et~al.},
  \bibinfo{journal}{Nat. Commun.} \textbf{\bibinfo{volume}{3}},
  \bibinfo{pages}{887} (\bibinfo{year}{2012}).

\bibitem[{\citenamefont{Mak et~al.}(2012)\citenamefont{Mak, He, Shan, and
  Heinz}}]{Mak}
\bibinfo{author}{\bibfnamefont{K.~F.} \bibnamefont{Mak}},
  \bibinfo{author}{\bibfnamefont{K.}~\bibnamefont{He}},
  \bibinfo{author}{\bibfnamefont{J.}~\bibnamefont{Shan}}, \bibnamefont{and}
  \bibinfo{author}{\bibfnamefont{T.~F.} \bibnamefont{Heinz}},
  \bibinfo{journal}{Nat. Nano.} \textbf{\bibinfo{volume}{7}},
  \bibinfo{pages}{494} (\bibinfo{year}{2012}).

\bibitem[{\citenamefont{Zeng et~al.}(2012)\citenamefont{Zeng, Dai, Yao, Xiao,
  and Cui}}]{Zeng}
\bibinfo{author}{\bibfnamefont{H.}~\bibnamefont{Zeng}},
  \bibinfo{author}{\bibfnamefont{J.}~\bibnamefont{Dai}},
  \bibinfo{author}{\bibfnamefont{W.}~\bibnamefont{Yao}},
  \bibinfo{author}{\bibfnamefont{D.}~\bibnamefont{Xiao}}, \bibnamefont{and}
  \bibinfo{author}{\bibfnamefont{X.}~\bibnamefont{Cui}}, \bibinfo{journal}{Nat.
  Nano.} \textbf{\bibinfo{volume}{7}}, \bibinfo{pages}{490}
  (\bibinfo{year}{2012}).

\bibitem[{\citenamefont{Isberg et~al.}(2013)\citenamefont{Isberg, Gabrysch,
  Hammersberg, Majdi, Kovi, and Twitchen}}]{Isberg}
\bibinfo{author}{\bibfnamefont{J.}~\bibnamefont{Isberg}},
  \bibinfo{author}{\bibfnamefont{M.}~\bibnamefont{Gabrysch}},
  \bibinfo{author}{\bibfnamefont{J.}~\bibnamefont{Hammersberg}},
  \bibinfo{author}{\bibfnamefont{S.}~\bibnamefont{Majdi}},
  \bibinfo{author}{\bibfnamefont{K.~K.} \bibnamefont{Kovi}}, \bibnamefont{and}
  \bibinfo{author}{\bibfnamefont{D.~J.} \bibnamefont{Twitchen}},
  \bibinfo{journal}{Nat. Mater.} \textbf{\bibinfo{volume}{12}},
  \bibinfo{pages}{760} (\bibinfo{year}{2013}).

\bibitem[{\citenamefont{Laird et~al.}(2013)\citenamefont{Laird, Pei, and
  Kouwenhoven}}]{Laird-nn}
\bibinfo{author}{\bibfnamefont{E.~A.} \bibnamefont{Laird}},
  \bibinfo{author}{\bibfnamefont{F.}~\bibnamefont{Pei}}, \bibnamefont{and}
  \bibinfo{author}{\bibfnamefont{L.~P.} \bibnamefont{Kouwenhoven}},
  \bibinfo{journal}{Nat. Nanotech.} \textbf{\bibinfo{volume}{8}},
  \bibinfo{pages}{565} (\bibinfo{year}{2013}).

\bibitem[{\citenamefont{Gorbachev et~al.}(2014)\citenamefont{Gorbachev, Song,
  Yu, Kretinin, Withers, Cao, Mishchenko, Grigorieva, Novoselov, Levitov
  et~al.}}]{GorbachevScience2014}
\bibinfo{author}{\bibfnamefont{R.~V.} \bibnamefont{Gorbachev}},
  \bibinfo{author}{\bibfnamefont{J.~C.~W.} \bibnamefont{Song}},
  \bibinfo{author}{\bibfnamefont{G.~L.} \bibnamefont{Yu}},
  \bibinfo{author}{\bibfnamefont{A.~V.} \bibnamefont{Kretinin}},
  \bibinfo{author}{\bibfnamefont{F.}~\bibnamefont{Withers}},
  \bibinfo{author}{\bibfnamefont{Y.}~\bibnamefont{Cao}},
  \bibinfo{author}{\bibfnamefont{A.}~\bibnamefont{Mishchenko}},
  \bibinfo{author}{\bibfnamefont{I.~V.} \bibnamefont{Grigorieva}},
  \bibinfo{author}{\bibfnamefont{K.~S.} \bibnamefont{Novoselov}},
  \bibinfo{author}{\bibfnamefont{L.~S.} \bibnamefont{Levitov}},
  \bibnamefont{et~al.}, \bibinfo{journal}{Science}
  \textbf{\bibinfo{volume}{346}}, \bibinfo{pages}{448} (\bibinfo{year}{2014}).

\bibitem[{\citenamefont{Mak et~al.}(2014)\citenamefont{Mak, McGill, Park, and
  McEuen}}]{MakScience2014}
\bibinfo{author}{\bibfnamefont{K.~F.} \bibnamefont{Mak}},
  \bibinfo{author}{\bibfnamefont{K.~L.} \bibnamefont{McGill}},
  \bibinfo{author}{\bibfnamefont{J.}~\bibnamefont{Park}}, \bibnamefont{and}
  \bibinfo{author}{\bibfnamefont{P.~L.} \bibnamefont{McEuen}},
  \bibinfo{journal}{Science} \textbf{\bibinfo{volume}{344}},
  \bibinfo{pages}{1489} (\bibinfo{year}{2014}).

\bibitem[{\citenamefont{Recher et~al.}(2007)\citenamefont{Recher, Trauzettel,
  Rycerz, Blanter, Beenakker, and Morpurgo}}]{Recher}
\bibinfo{author}{\bibfnamefont{P.}~\bibnamefont{Recher}},
  \bibinfo{author}{\bibfnamefont{B.}~\bibnamefont{Trauzettel}},
  \bibinfo{author}{\bibfnamefont{A.}~\bibnamefont{Rycerz}},
  \bibinfo{author}{\bibfnamefont{Y.~M.} \bibnamefont{Blanter}},
  \bibinfo{author}{\bibfnamefont{C.~W.~J.} \bibnamefont{Beenakker}},
  \bibnamefont{and} \bibinfo{author}{\bibfnamefont{A.~F.}
  \bibnamefont{Morpurgo}}, \bibinfo{journal}{Phys. Rev. B}
  \textbf{\bibinfo{volume}{76}}, \bibinfo{pages}{235404}
  (\bibinfo{year}{2007}).

\bibitem[{\citenamefont{P\'alyi and Burkard}(2011)}]{Palyi-valley-resonance}
\bibinfo{author}{\bibfnamefont{A.}~\bibnamefont{P\'alyi}} \bibnamefont{and}
  \bibinfo{author}{\bibfnamefont{G.}~\bibnamefont{Burkard}},
  \bibinfo{journal}{Phys. Rev. Lett.} \textbf{\bibinfo{volume}{106}},
  \bibinfo{pages}{086801} (\bibinfo{year}{2011}).

\bibitem[{\citenamefont{Wu et~al.}(2011)\citenamefont{Wu, Lue, and
  Chang}}]{WuPRB2011}
\bibinfo{author}{\bibfnamefont{G.~Y.} \bibnamefont{Wu}},
  \bibinfo{author}{\bibfnamefont{N.-Y.} \bibnamefont{Lue}}, \bibnamefont{and}
  \bibinfo{author}{\bibfnamefont{L.}~\bibnamefont{Chang}},
  \bibinfo{journal}{Phys. Rev. B} \textbf{\bibinfo{volume}{84}},
  \bibinfo{pages}{195463} (\bibinfo{year}{2011}).

\bibitem[{\citenamefont{Wu and Lue}(2012)}]{WuPRB2012}
\bibinfo{author}{\bibfnamefont{G.~Y.} \bibnamefont{Wu}} \bibnamefont{and}
  \bibinfo{author}{\bibfnamefont{N.-Y.} \bibnamefont{Lue}},
  \bibinfo{journal}{Phys. Rev. B} \textbf{\bibinfo{volume}{86}},
  \bibinfo{pages}{045456} (\bibinfo{year}{2012}).

\bibitem[{\citenamefont{Wu et~al.}(2013)\citenamefont{Wu, Lue, and
  Chen}}]{WuPRB2013}
\bibinfo{author}{\bibfnamefont{G.~Y.} \bibnamefont{Wu}},
  \bibinfo{author}{\bibfnamefont{N.-Y.} \bibnamefont{Lue}}, \bibnamefont{and}
  \bibinfo{author}{\bibfnamefont{Y.-C.} \bibnamefont{Chen}},
  \bibinfo{journal}{Phys. Rev. B} \textbf{\bibinfo{volume}{88}},
  \bibinfo{pages}{125422} (\bibinfo{year}{2013}).

\bibitem[{\citenamefont{Culcer et~al.}(2012)\citenamefont{Culcer, Saraiva,
  Koiller, Hu, and Das~Sarma}}]{CulcerPRL}
\bibinfo{author}{\bibfnamefont{D.}~\bibnamefont{Culcer}},
  \bibinfo{author}{\bibfnamefont{A.~L.} \bibnamefont{Saraiva}},
  \bibinfo{author}{\bibfnamefont{B.}~\bibnamefont{Koiller}},
  \bibinfo{author}{\bibfnamefont{X.}~\bibnamefont{Hu}}, \bibnamefont{and}
  \bibinfo{author}{\bibfnamefont{S.}~\bibnamefont{Das~Sarma}},
  \bibinfo{journal}{Phys. Rev. Lett.} \textbf{\bibinfo{volume}{108}},
  \bibinfo{pages}{126804} (\bibinfo{year}{2012}).

\bibitem[{\citenamefont{Sz\'echenyi and P\'alyi}(2014)}]{Szechenyi-maximalrabi}
\bibinfo{author}{\bibfnamefont{G.}~\bibnamefont{Sz\'echenyi}} \bibnamefont{and}
  \bibinfo{author}{\bibfnamefont{A.}~\bibnamefont{P\'alyi}},
  \bibinfo{journal}{Phys. Rev. B} \textbf{\bibinfo{volume}{89}},
  \bibinfo{pages}{115409} (\bibinfo{year}{2014}).

\bibitem[{\citenamefont{Korm\'anyos et~al.}(2014)\citenamefont{Korm\'anyos,
  Z\'olyomi, Drummond, and Burkard}}]{KormanyosPRX2014}
\bibinfo{author}{\bibfnamefont{A.}~\bibnamefont{Korm\'anyos}},
  \bibinfo{author}{\bibfnamefont{V.}~\bibnamefont{Z\'olyomi}},
  \bibinfo{author}{\bibfnamefont{N.~D.} \bibnamefont{Drummond}},
  \bibnamefont{and} \bibinfo{author}{\bibfnamefont{G.}~\bibnamefont{Burkard}},
  \bibinfo{journal}{Phys. Rev. X} \textbf{\bibinfo{volume}{4}},
  \bibinfo{pages}{011034} (\bibinfo{year}{2014}).

\bibitem[{\citenamefont{Liu et~al.}(2014)\citenamefont{Liu, Pang, Yao, and
  Yao}}]{GuiBinLiuNJP2014}
\bibinfo{author}{\bibfnamefont{G.-B.} \bibnamefont{Liu}},
  \bibinfo{author}{\bibfnamefont{H.}~\bibnamefont{Pang}},
  \bibinfo{author}{\bibfnamefont{Y.}~\bibnamefont{Yao}}, \bibnamefont{and}
  \bibinfo{author}{\bibfnamefont{W.}~\bibnamefont{Yao}}, \bibinfo{journal}{New
  Journal of Physics} \textbf{\bibinfo{volume}{16}}, \bibinfo{pages}{105011}
  (\bibinfo{year}{2014}).

\bibitem[{\citenamefont{Rohling and Burkard}(2012)}]{Rohling-njp}
\bibinfo{author}{\bibfnamefont{N.}~\bibnamefont{Rohling}} \bibnamefont{and}
  \bibinfo{author}{\bibfnamefont{G.}~\bibnamefont{Burkard}},
  \bibinfo{journal}{New Journal of Physics} \textbf{\bibinfo{volume}{14}},
  \bibinfo{pages}{083008} (\bibinfo{year}{2012}).

\bibitem[{\citenamefont{Rohling et~al.}(2014)\citenamefont{Rohling, Russ, and
  Burkard}}]{Rohling-prl}
\bibinfo{author}{\bibfnamefont{N.}~\bibnamefont{Rohling}},
  \bibinfo{author}{\bibfnamefont{M.}~\bibnamefont{Russ}}, \bibnamefont{and}
  \bibinfo{author}{\bibfnamefont{G.}~\bibnamefont{Burkard}},
  \bibinfo{journal}{Phys. Rev. Lett.} \textbf{\bibinfo{volume}{113}},
  \bibinfo{pages}{176801} (\bibinfo{year}{2014}).

\bibitem[{\citenamefont{Geim and Novoselov}(2007)}]{Geim-rise}
\bibinfo{author}{\bibfnamefont{A.~K.} \bibnamefont{Geim}} \bibnamefont{and}
  \bibinfo{author}{\bibfnamefont{K.~S.} \bibnamefont{Novoselov}},
  \bibinfo{journal}{Nat. Mater.} \textbf{\bibinfo{volume}{6}},
  \bibinfo{pages}{183} (\bibinfo{year}{2007}).

\bibitem[{\citenamefont{Castro~Neto et~al.}(2009)\citenamefont{Castro~Neto,
  Guinea, Peres, Novoselov, and Geim}}]{CastroNeto-rmp}
\bibinfo{author}{\bibfnamefont{A.~H.} \bibnamefont{Castro~Neto}},
  \bibinfo{author}{\bibfnamefont{F.}~\bibnamefont{Guinea}},
  \bibinfo{author}{\bibfnamefont{N.~M.~R.} \bibnamefont{Peres}},
  \bibinfo{author}{\bibfnamefont{K.~S.} \bibnamefont{Novoselov}},
  \bibnamefont{and} \bibinfo{author}{\bibfnamefont{A.~K.} \bibnamefont{Geim}},
  \bibinfo{journal}{Rev. Mod. Phys.} \textbf{\bibinfo{volume}{81}},
  \bibinfo{pages}{109} (\bibinfo{year}{2009}).

\bibitem[{\citenamefont{Ando}(2006)}]{AndoJPSJ}
\bibinfo{author}{\bibfnamefont{T.}~\bibnamefont{Ando}},
  \bibinfo{journal}{Journal of the Physical Society of Japan}
  \textbf{\bibinfo{volume}{75}}, \bibinfo{pages}{074716}
  (\bibinfo{year}{2006}).

\bibitem[{\citenamefont{Hwang et~al.}(2007)\citenamefont{Hwang, Adam, and
  Sarma}}]{HwangPRB-transport}
\bibinfo{author}{\bibfnamefont{E.~H.} \bibnamefont{Hwang}},
  \bibinfo{author}{\bibfnamefont{S.}~\bibnamefont{Adam}}, \bibnamefont{and}
  \bibinfo{author}{\bibfnamefont{S.~D.} \bibnamefont{Sarma}},
  \bibinfo{journal}{Phys. Rev. Lett.} \textbf{\bibinfo{volume}{98}},
  \bibinfo{pages}{186806} (\bibinfo{year}{2007}).

\bibitem[{\citenamefont{Song and Dery}(2013)}]{Song}
\bibinfo{author}{\bibfnamefont{Y.}~\bibnamefont{Song}} \bibnamefont{and}
  \bibinfo{author}{\bibfnamefont{H.}~\bibnamefont{Dery}},
  \bibinfo{journal}{Phys. Rev. Lett.} \textbf{\bibinfo{volume}{111}},
  \bibinfo{pages}{026601} (\bibinfo{year}{2013}).

\bibitem[{\citenamefont{Ochoa et~al.}(2014)\citenamefont{Ochoa, Finocchiaro,
  Guinea, and Fal'ko}}]{Ochoa}
\bibinfo{author}{\bibfnamefont{H.}~\bibnamefont{Ochoa}},
  \bibinfo{author}{\bibfnamefont{F.}~\bibnamefont{Finocchiaro}},
  \bibinfo{author}{\bibfnamefont{F.}~\bibnamefont{Guinea}}, \bibnamefont{and}
  \bibinfo{author}{\bibfnamefont{V.~I.} \bibnamefont{Fal'ko}},
  \bibinfo{journal}{Phys. Rev. B} \textbf{\bibinfo{volume}{90}},
  \bibinfo{pages}{235429} (\bibinfo{year}{2014}).

\bibitem[{\citenamefont{Mai et~al.}(2014)\citenamefont{Mai, Semenov, Barrette,
  Yu, Jin, Cao, Kim, and Gundogdu}}]{CongMaiPRB2014}
\bibinfo{author}{\bibfnamefont{C.}~\bibnamefont{Mai}},
  \bibinfo{author}{\bibfnamefont{Y.~G.} \bibnamefont{Semenov}},
  \bibinfo{author}{\bibfnamefont{A.}~\bibnamefont{Barrette}},
  \bibinfo{author}{\bibfnamefont{Y.}~\bibnamefont{Yu}},
  \bibinfo{author}{\bibfnamefont{Z.}~\bibnamefont{Jin}},
  \bibinfo{author}{\bibfnamefont{L.}~\bibnamefont{Cao}},
  \bibinfo{author}{\bibfnamefont{K.~W.} \bibnamefont{Kim}}, \bibnamefont{and}
  \bibinfo{author}{\bibfnamefont{K.}~\bibnamefont{Gundogdu}},
  \bibinfo{journal}{Phys. Rev. B} \textbf{\bibinfo{volume}{90}},
  \bibinfo{pages}{041414} (\bibinfo{year}{2014}).

\bibitem[{\citenamefont{Pachoud et~al.}(2014)\citenamefont{Pachoud, Ferreira,
  \"Ozyilmaz, and Castro~Neto}}]{Pachoud}
\bibinfo{author}{\bibfnamefont{A.}~\bibnamefont{Pachoud}},
  \bibinfo{author}{\bibfnamefont{A.}~\bibnamefont{Ferreira}},
  \bibinfo{author}{\bibfnamefont{B.}~\bibnamefont{\"Ozyilmaz}},
  \bibnamefont{and} \bibinfo{author}{\bibfnamefont{A.~H.}
  \bibnamefont{Castro~Neto}}, \bibinfo{journal}{Phys. Rev. B}
  \textbf{\bibinfo{volume}{90}}, \bibinfo{pages}{035444}
  (\bibinfo{year}{2014}).

\bibitem[{\citenamefont{Braginsky and Entin}()}]{Braginsky}
\bibinfo{author}{\bibfnamefont{L.}~\bibnamefont{Braginsky}} \bibnamefont{and}
  \bibinfo{author}{\bibfnamefont{M.}~\bibnamefont{Entin}},
  \bibinfo{note}{arXiv:1412.7810 (unpublished)}.

\bibitem[{\citenamefont{Csisz\'ar and P\'alyi}(2014)}]{Csiszar}
\bibinfo{author}{\bibfnamefont{G.}~\bibnamefont{Csisz\'ar}} \bibnamefont{and}
  \bibinfo{author}{\bibfnamefont{A.}~\bibnamefont{P\'alyi}},
  \bibinfo{journal}{Phys. Rev. B} \textbf{\bibinfo{volume}{90}},
  \bibinfo{pages}{245413} (\bibinfo{year}{2014}).

\bibitem[{\citenamefont{Tahan and Joynt}(2014)}]{Tahan}
\bibinfo{author}{\bibfnamefont{C.}~\bibnamefont{Tahan}} \bibnamefont{and}
  \bibinfo{author}{\bibfnamefont{R.}~\bibnamefont{Joynt}},
  \bibinfo{journal}{Phys. Rev. B} \textbf{\bibinfo{volume}{89}},
  \bibinfo{pages}{075302} (\bibinfo{year}{2014}).

\bibitem[{\citenamefont{Yang et~al.}(2013)\citenamefont{Yang, Rossi, Ruskov,
  Lai, Mohiyaddin, Lee, Tahan, Klimeck, Morello, and Dzurak}}]{Yang}
\bibinfo{author}{\bibfnamefont{C.~H.} \bibnamefont{Yang}},
  \bibinfo{author}{\bibfnamefont{A.}~\bibnamefont{Rossi}},
  \bibinfo{author}{\bibfnamefont{R.}~\bibnamefont{Ruskov}},
  \bibinfo{author}{\bibfnamefont{N.~S.} \bibnamefont{Lai}},
  \bibinfo{author}{\bibfnamefont{F.~A.} \bibnamefont{Mohiyaddin}},
  \bibinfo{author}{\bibfnamefont{S.}~\bibnamefont{Lee}},
  \bibinfo{author}{\bibfnamefont{C.}~\bibnamefont{Tahan}},
  \bibinfo{author}{\bibfnamefont{G.}~\bibnamefont{Klimeck}},
  \bibinfo{author}{\bibfnamefont{A.}~\bibnamefont{Morello}}, \bibnamefont{and}
  \bibinfo{author}{\bibfnamefont{A.~S.} \bibnamefont{Dzurak}},
  \bibinfo{journal}{Nat. Commun.} \textbf{\bibinfo{volume}{4}}
  (\bibinfo{year}{2013}).

\bibitem[{\citenamefont{Adler}(1962)}]{AdlerPR}
\bibinfo{author}{\bibfnamefont{S.~L.} \bibnamefont{Adler}},
  \bibinfo{journal}{Phys. Rev.} \textbf{\bibinfo{volume}{126}},
  \bibinfo{pages}{413} (\bibinfo{year}{1962}).

\bibitem[{\citenamefont{Wiser}(1963)}]{WiserPR}
\bibinfo{author}{\bibfnamefont{N.}~\bibnamefont{Wiser}},
  \bibinfo{journal}{Phys. Rev.} \textbf{\bibinfo{volume}{129}},
  \bibinfo{pages}{62} (\bibinfo{year}{1963}).

\bibitem[{\citenamefont{van Schilfgaarde and Katsnelson}(2011)}]{Schilfgaarde}
\bibinfo{author}{\bibfnamefont{M.}~\bibnamefont{van Schilfgaarde}}
  \bibnamefont{and} \bibinfo{author}{\bibfnamefont{M.~I.}
  \bibnamefont{Katsnelson}}, \bibinfo{journal}{Phys. Rev. B}
  \textbf{\bibinfo{volume}{83}}, \bibinfo{pages}{081409}
  (\bibinfo{year}{2011}).

\bibitem[{\citenamefont{Tudorovskiy and Mikhailov}(2010)}]{Tudorovskiy}
\bibinfo{author}{\bibfnamefont{T.}~\bibnamefont{Tudorovskiy}} \bibnamefont{and}
  \bibinfo{author}{\bibfnamefont{S.~A.} \bibnamefont{Mikhailov}},
  \bibinfo{journal}{Phys. Rev. B} \textbf{\bibinfo{volume}{82}},
  \bibinfo{pages}{073411} (\bibinfo{year}{2010}).

\bibitem[{\citenamefont{Suzuura and Ando}(2002)}]{Suzuura}
\bibinfo{author}{\bibfnamefont{H.}~\bibnamefont{Suzuura}} \bibnamefont{and}
  \bibinfo{author}{\bibfnamefont{T.}~\bibnamefont{Ando}},
  \bibinfo{journal}{Phys. Rev. Lett.} \textbf{\bibinfo{volume}{89}},
  \bibinfo{pages}{266603} (\bibinfo{year}{2002}).

\bibitem[{\citenamefont{McCann et~al.}(2006)\citenamefont{McCann, Kechedzhi,
  Fal'ko, Suzuura, Ando, and Altshuler}}]{McCannPRL}
\bibinfo{author}{\bibfnamefont{E.}~\bibnamefont{McCann}},
  \bibinfo{author}{\bibfnamefont{K.}~\bibnamefont{Kechedzhi}},
  \bibinfo{author}{\bibfnamefont{V.~I.} \bibnamefont{Fal'ko}},
  \bibinfo{author}{\bibfnamefont{H.}~\bibnamefont{Suzuura}},
  \bibinfo{author}{\bibfnamefont{T.}~\bibnamefont{Ando}}, \bibnamefont{and}
  \bibinfo{author}{\bibfnamefont{B.~L.} \bibnamefont{Altshuler}},
  \bibinfo{journal}{Phys. Rev. Lett.} \textbf{\bibinfo{volume}{97}},
  \bibinfo{pages}{146805} (\bibinfo{year}{2006}).

\bibitem[{\citenamefont{Morozov et~al.}(2006)\citenamefont{Morozov, Novoselov,
  Katsnelson, Schedin, Ponomarenko, Jiang, and Geim}}]{MorozovPRL}
\bibinfo{author}{\bibfnamefont{S.~V.} \bibnamefont{Morozov}},
  \bibinfo{author}{\bibfnamefont{K.~S.} \bibnamefont{Novoselov}},
  \bibinfo{author}{\bibfnamefont{M.~I.} \bibnamefont{Katsnelson}},
  \bibinfo{author}{\bibfnamefont{F.}~\bibnamefont{Schedin}},
  \bibinfo{author}{\bibfnamefont{L.~A.} \bibnamefont{Ponomarenko}},
  \bibinfo{author}{\bibfnamefont{D.}~\bibnamefont{Jiang}}, \bibnamefont{and}
  \bibinfo{author}{\bibfnamefont{A.~K.} \bibnamefont{Geim}},
  \bibinfo{journal}{Phys. Rev. Lett.} \textbf{\bibinfo{volume}{97}},
  \bibinfo{pages}{016801} (\bibinfo{year}{2006}).

\bibitem[{\citenamefont{Wu et~al.}(2007)\citenamefont{Wu, Li, Song, Berger, and
  de~Heer}}]{WuPRL2007}
\bibinfo{author}{\bibfnamefont{X.}~\bibnamefont{Wu}},
  \bibinfo{author}{\bibfnamefont{X.}~\bibnamefont{Li}},
  \bibinfo{author}{\bibfnamefont{Z.}~\bibnamefont{Song}},
  \bibinfo{author}{\bibfnamefont{C.}~\bibnamefont{Berger}}, \bibnamefont{and}
  \bibinfo{author}{\bibfnamefont{W.~A.} \bibnamefont{de~Heer}},
  \bibinfo{journal}{Phys. Rev. Lett.} \textbf{\bibinfo{volume}{98}},
  \bibinfo{pages}{136801} (\bibinfo{year}{2007}).

\bibitem[{\citenamefont{Tikhonenko et~al.}(2008)\citenamefont{Tikhonenko,
  Horsell, Gorbachev, and Savchenko}}]{TikhonenkoPRL2008}
\bibinfo{author}{\bibfnamefont{F.~V.} \bibnamefont{Tikhonenko}},
  \bibinfo{author}{\bibfnamefont{D.~W.} \bibnamefont{Horsell}},
  \bibinfo{author}{\bibfnamefont{R.~V.} \bibnamefont{Gorbachev}},
  \bibnamefont{and} \bibinfo{author}{\bibfnamefont{A.~K.}
  \bibnamefont{Savchenko}}, \bibinfo{journal}{Phys. Rev. Lett.}
  \textbf{\bibinfo{volume}{100}}, \bibinfo{pages}{056802}
  (\bibinfo{year}{2008}).

\bibitem[{\citenamefont{Malard et~al.}(2009)\citenamefont{Malard, Pimenta,
  Dresselhaus, and Dresselhaus}}]{MalardPR2009}
\bibinfo{author}{\bibfnamefont{L.}~\bibnamefont{Malard}},
  \bibinfo{author}{\bibfnamefont{M.}~\bibnamefont{Pimenta}},
  \bibinfo{author}{\bibfnamefont{G.}~\bibnamefont{Dresselhaus}},
  \bibnamefont{and}
  \bibinfo{author}{\bibfnamefont{M.}~\bibnamefont{Dresselhaus}},
  \bibinfo{journal}{Physics Reports} \textbf{\bibinfo{volume}{473}},
  \bibinfo{pages}{51 } (\bibinfo{year}{2009}).

\bibitem[{\citenamefont{Graf et~al.}(2007)\citenamefont{Graf, Molitor, Ensslin,
  Stampfer, Jungen, Hierold, and Wirtz}}]{GrafNanoLett2007}
\bibinfo{author}{\bibfnamefont{D.}~\bibnamefont{Graf}},
  \bibinfo{author}{\bibfnamefont{F.}~\bibnamefont{Molitor}},
  \bibinfo{author}{\bibfnamefont{K.}~\bibnamefont{Ensslin}},
  \bibinfo{author}{\bibfnamefont{C.}~\bibnamefont{Stampfer}},
  \bibinfo{author}{\bibfnamefont{A.}~\bibnamefont{Jungen}},
  \bibinfo{author}{\bibfnamefont{C.}~\bibnamefont{Hierold}}, \bibnamefont{and}
  \bibinfo{author}{\bibfnamefont{L.}~\bibnamefont{Wirtz}},
  \bibinfo{journal}{Nano Letters} \textbf{\bibinfo{volume}{7}},
  \bibinfo{pages}{238} (\bibinfo{year}{2007}).

\bibitem[{\citenamefont{Wallace}(1947)}]{WallacePR}
\bibinfo{author}{\bibfnamefont{P.~R.} \bibnamefont{Wallace}},
  \bibinfo{journal}{Phys. Rev.} \textbf{\bibinfo{volume}{71}},
  \bibinfo{pages}{622} (\bibinfo{year}{1947}).

\bibitem[{\citenamefont{Das~Sarma et~al.}(2011)\citenamefont{Das~Sarma, Adam,
  Hwang, and Rossi}}]{DasSarma-graphenereview}
\bibinfo{author}{\bibfnamefont{S.}~\bibnamefont{Das~Sarma}},
  \bibinfo{author}{\bibfnamefont{S.}~\bibnamefont{Adam}},
  \bibinfo{author}{\bibfnamefont{E.~H.} \bibnamefont{Hwang}}, \bibnamefont{and}
  \bibinfo{author}{\bibfnamefont{E.}~\bibnamefont{Rossi}},
  \bibinfo{journal}{Rev. Mod. Phys.} \textbf{\bibinfo{volume}{83}},
  \bibinfo{pages}{407} (\bibinfo{year}{2011}).

\bibitem[{\citenamefont{Fradkin}(1986{\natexlab{a}})}]{Fradkin1}
\bibinfo{author}{\bibfnamefont{E.}~\bibnamefont{Fradkin}},
  \bibinfo{journal}{Phys. Rev. B} \textbf{\bibinfo{volume}{33}},
  \bibinfo{pages}{3257} (\bibinfo{year}{1986}{\natexlab{a}}).

\bibitem[{\citenamefont{Fradkin}(1986{\natexlab{b}})}]{Fradkin2}
\bibinfo{author}{\bibfnamefont{E.}~\bibnamefont{Fradkin}},
  \bibinfo{journal}{Phys. Rev. B} \textbf{\bibinfo{volume}{33}},
  \bibinfo{pages}{3263} (\bibinfo{year}{1986}{\natexlab{b}}).

\bibitem[{\citenamefont{Hwang and Das~Sarma}(2007)}]{HwangPRB-dielectric}
\bibinfo{author}{\bibfnamefont{E.~H.} \bibnamefont{Hwang}} \bibnamefont{and}
  \bibinfo{author}{\bibfnamefont{S.}~\bibnamefont{Das~Sarma}},
  \bibinfo{journal}{Phys. Rev. B} \textbf{\bibinfo{volume}{75}},
  \bibinfo{pages}{205418} (\bibinfo{year}{2007}).

\bibitem[{\citenamefont{Wunsch et~al.}(2006)\citenamefont{Wunsch, Stauber,
  Sols, and Guinea}}]{WunschNJP}
\bibinfo{author}{\bibfnamefont{B.}~\bibnamefont{Wunsch}},
  \bibinfo{author}{\bibfnamefont{T.}~\bibnamefont{Stauber}},
  \bibinfo{author}{\bibfnamefont{F.}~\bibnamefont{Sols}}, \bibnamefont{and}
  \bibinfo{author}{\bibfnamefont{F.}~\bibnamefont{Guinea}},
  \bibinfo{journal}{New Journal of Physics} \textbf{\bibinfo{volume}{8}},
  \bibinfo{pages}{318} (\bibinfo{year}{2006}).

\bibitem[{\citenamefont{Lindhard and Winther}(1964)}]{Lindhard}
\bibinfo{author}{\bibfnamefont{J.}~\bibnamefont{Lindhard}} \bibnamefont{and}
  \bibinfo{author}{\bibfnamefont{A.}~\bibnamefont{Winther}},
  \bibinfo{journal}{Mat. Fys. Medd. Dan. Vid. Selsk.}
  \textbf{\bibinfo{volume}{34}}, \bibinfo{pages}{4} (\bibinfo{year}{1964}).

\bibitem[{\citenamefont{Pereira et~al.}(2007)\citenamefont{Pereira, Nilsson,
  and Castro~Neto}}]{Pereira}
\bibinfo{author}{\bibfnamefont{V.~M.} \bibnamefont{Pereira}},
  \bibinfo{author}{\bibfnamefont{J.}~\bibnamefont{Nilsson}}, \bibnamefont{and}
  \bibinfo{author}{\bibfnamefont{A.~H.} \bibnamefont{Castro~Neto}},
  \bibinfo{journal}{Phys. Rev. Lett.} \textbf{\bibinfo{volume}{99}},
  \bibinfo{pages}{166802} (\bibinfo{year}{2007}).

\bibitem[{\citenamefont{Shytov et~al.}(2007)\citenamefont{Shytov, Katsnelson,
  and Levitov}}]{Shytov}
\bibinfo{author}{\bibfnamefont{A.~V.} \bibnamefont{Shytov}},
  \bibinfo{author}{\bibfnamefont{M.~I.} \bibnamefont{Katsnelson}},
  \bibnamefont{and} \bibinfo{author}{\bibfnamefont{L.~S.}
  \bibnamefont{Levitov}}, \bibinfo{journal}{Phys. Rev. Lett.}
  \textbf{\bibinfo{volume}{99}}, \bibinfo{pages}{236801}
  (\bibinfo{year}{2007}).

\bibitem[{\citenamefont{Novikov}(2007)}]{Novikov}
\bibinfo{author}{\bibfnamefont{D.~S.} \bibnamefont{Novikov}},
  \bibinfo{journal}{Phys. Rev. B} \textbf{\bibinfo{volume}{76}},
  \bibinfo{pages}{245435} (\bibinfo{year}{2007}).

\bibitem[{\citenamefont{Valet and Fert}(1993)}]{Valet}
\bibinfo{author}{\bibfnamefont{T.}~\bibnamefont{Valet}} \bibnamefont{and}
  \bibinfo{author}{\bibfnamefont{A.}~\bibnamefont{Fert}},
  \bibinfo{journal}{Phys. Rev. B} \textbf{\bibinfo{volume}{48}},
  \bibinfo{pages}{7099} (\bibinfo{year}{1993}).

\bibitem[{\citenamefont{Abanin et~al.}(2009)\citenamefont{Abanin, Shytov,
  Levitov, and Halperin}}]{AbaninPRB2009}
\bibinfo{author}{\bibfnamefont{D.~A.} \bibnamefont{Abanin}},
  \bibinfo{author}{\bibfnamefont{A.~V.} \bibnamefont{Shytov}},
  \bibinfo{author}{\bibfnamefont{L.~S.} \bibnamefont{Levitov}},
  \bibnamefont{and} \bibinfo{author}{\bibfnamefont{B.~I.}
  \bibnamefont{Halperin}}, \bibinfo{journal}{Phys. Rev. B}
  \textbf{\bibinfo{volume}{79}}, \bibinfo{pages}{035304}
  (\bibinfo{year}{2009}).

\bibitem[{\citenamefont{Clementi and Raimondi}(1963)}]{ClementiRaimondi}
\bibinfo{author}{\bibfnamefont{E.}~\bibnamefont{Clementi}} \bibnamefont{and}
  \bibinfo{author}{\bibfnamefont{D.~L.} \bibnamefont{Raimondi}},
  \bibinfo{journal}{The Journal of Chemical Physics}
  \textbf{\bibinfo{volume}{38}}, \bibinfo{pages}{2686} (\bibinfo{year}{1963}).

\end{thebibliography}
\end{document}